\documentclass[prx,longbibliography,twocolumn,showpacs,nofootinbib,superscriptaddress,notitlepage]{revtex4-1}
\usepackage{amsmath}
\usepackage{amssymb,bm}
\usepackage{amsthm}
\usepackage{mathtools}
\usepackage{url}
\DeclarePairedDelimiter\floor{\lfloor}{\rfloor}

\newtheorem{innercustomgeneric}{\customgenericname}
\providecommand{\customgenericname}{}
\newcommand{\newcustomtheorem}[2]{%
  \newenvironment{#1}[1]
  {%
   \renewcommand\customgenericname{#2}%
   \renewcommand\theinnercustomgeneric{##1}%
   \innercustomgeneric
  }
  {\endinnercustomgeneric}
}

\newcustomtheorem{customthm}{Theorem}
\newcustomtheorem{definitionBob}{Definition}

\usepackage{color,dsfont}
\usepackage{graphicx}
\usepackage{ragged2e}
\usepackage[colorlinks=true, hyperindex, breaklinks, linkcolor=blue, urlcolor=blue, citecolor=blue]{hyperref} 
\usepackage[normalem]{ulem}
\usepackage[capitalise]{cleveref}
\usepackage{mathrsfs}
\usepackage[caption=false]{subfig}
\usepackage{mathtools}
\usepackage{float}
\usepackage{verbatim}
\usepackage{latexsym}
\usepackage{amsmath}
\usepackage{amssymb}
\usepackage{setspace}
\usepackage{amsfonts}
\usepackage{stmaryrd}
\usepackage{xcolor}
\usepackage{enumitem}
\usepackage{lipsum}

\newtheorem{theorem}{Theorem}

\newcommand{\ket}[1]{|#1\rangle}  

\newcommand{\codepar}[1]{\ensuremath{[\![#1]\!]}}

\newcommand{\License}[1]{\begingroup
  \renewcommand\thefootnote{}
  \footnote{#1}%
  \addtocounter{footnote}{-1}%
  \endgroup
}

\begin{document}

\title{Topological and subsystem codes on low-degree graphs with flag qubits}

\author{Christopher Chamberland}
\email{mathematicschris@gmail.com}
\affiliation{
   IBM T.~J.~Watson Research Center,
    Yorktown Heights, NY, 10598, United States
    }

 \author{Guanyu Zhu}
\email{guanyu.zhu@ibm.com}
\affiliation{
   IBM T.~J.~Watson Research Center,
    Yorktown Heights, NY, 10598, United States
    }

 \author{Theodore J. Yoder}
\affiliation{
   IBM T.~J.~Watson Research Center,
    Yorktown Heights, NY, 10598, United States
}

    \author{Jared B. Hertzberg }
\affiliation{
   IBM T.~J.~Watson Research Center,
    Yorktown Heights, NY, 10598, United States
}

\author{Andrew W. Cross}
\affiliation{
   IBM T.~J.~Watson Research Center,
    Yorktown Heights, NY, 10598, United States
}
\License{C.C. and G.Z. were the main contributors of this work.}

\begin{abstract}

In this work we introduce two code families, which we call the heavy hexagon code and heavy square code. Both code families are implemented by assigning physical data and ancilla qubits to both vertices and edges of low degree graphs. Such a layout is particularly suitable for superconducting qubit architectures to minimize frequency collisions and crosstalk. In some cases, frequency collisions can be reduced by several orders of magnitude. The heavy hexagon code is a hybrid surface/Bacon-Shor code mapped onto a (heavy) hexagonal lattice whereas the heavy square code is the surface code mapped onto a (heavy) square lattice. In both cases, the lattice includes all the ancilla qubits required for fault-tolerant error-correction. Naively, the limited qubit connectivity might be thought to limit the error-correcting capability of the code to less than its full distance. Therefore, essential to our construction is the use of flag qubits. We modify minimum weight perfect matching decoding to efficiently and scalably incorporate information from measurements of the flag qubits and correct up to the full code distance while respecting the limited connectivity. Simulations show that high threshold values for both codes can be obtained using our decoding protocol. Further, our decoding scheme can be adapted to other topological code families.

\end{abstract}

\pacs{03.67.Pp}

\maketitle

\section{Introduction}
\label{sec:Intro}

Fault-tolerant quantum computing with quantum error correcting codes (QECC) \cite{DKLP02, FMMC12, Barbara15} is a scalable way to achieve universal quantum computation which will be capable of performing quantum algorithms that offer significant advantages over classical algorithms.  With the rapid development of quantum computing platforms such as superconducting circuits and ion traps in the past decade, the path towards achieving logical qubits with $\mathcal{O}(100)$ physical qubits and demonstrating fault tolerance in near term devices looks very promising.

Leading candidates of QECC in the near term include topological stabilizer codes such as the surface code \cite{DKLP02, FMMC12} and subsystem codes such as the Bacon-Shor code \cite{Bacon:2006ul, Aliferis:2007dx}. These codes belong to the class of quantum low-density-parity-check (LDPC) codes, and hence error-correction consists of measuring low-weight Pauli operators whose size is independent of the code distance. The standard schemes to implement these codes typically choose a square lattice which is motivated by minimizing the depth of the syndrome measurement circuits while allowing syndrome measurements to be performed using nearest neighbor interactions.

For implementations with superconducting circuits, promising architectures include fixed-frequency transmon qubits coupled via the cross resonance (CR) gates \cite{Rigetti2010, Chow2011}, tunable-frequency transmons coupled via the controlled-phase gate \cite{BKM+14, DiCarlo:2009ja},  systems using tunable couplers \cite{Chen:2014cw, McKay2016} and so on. In the context of the CR gates, the relative stability of microwave control as opposed to flux drive/tuning results in high fidelity gates which have achieved error rates below $1\%$ \cite{Sheldon2016} and hence approaching the surface code error threshold. Demonstrations of syndrome measurements and fault-tolerant protocols using post-selection in small scale devices has also been achieved \cite{Chow2014, Corcoles2015, Takita2016, Takita2017}. However, to implement the standard surface code within this architecture requires data and syndrome measurement qubits placed on a square lattice, where each vertex has degree four (with four neighboring qubits). Therefore, a minimum of five distinct frequencies is required for the experimental implementation to ensure individual addressability of the CR gates and the avoidance of crosstalk \cite{Gambetta2017}.  This imposes a significant challenge to the device fabrication process which has to avoid possible frequency collisions limiting the code performance.  Similar problems of crosstalk also exist in other superconducting architectures such as those using the controlled-phase gates.

In this paper, we design codes on low-degree graphs which can minimize the possibility of frequency collisions and optimize the hardware performance within superconducting qubit architectures. In particular, we have designed a family of subsystem codes on a ``heavy hexagon" lattice with a mixture of degree-two and degree-three vertices, which can be considered as a hybrid surface/Bacon-Shor code,  and a family of modified surface codes on a ``heavy square" lattice with a mixture of degree two and four vertices.  These codes reduce the distinct number of frequencies to only three in the bulk.  The price of reducing the degree is to introduce more ancilla qubits mediating the entanglement for the syndrome measurement, which results in the increase of the depth of the syndrome extraction circuits and hence potentially increases the logical error rate.  On the other hand, the extra ancillas can also become resources for the decoding process.  In particular, we have designed a protocol using the ancillas as flag qubits \cite{CR17v1,CR17v2,CB17,TCD18Flag,ReichardtFlag18,ChamberlandMagic,ChamberlandGKP}, which allows errors to be corrected up to the full code distance and hence significantly suppresses the logical error rate (thoughout the manuscript, unless specified otherwise, the term ancilla will be used for both syndrome measurement and flag qubits).  When implementing the flag decoder, the heavy square code can achieve an error threshold of approximately $0.3 \%$ for both $X$ and $Z$ errors, while the heavy hexagon code achieves a threshold of approximately $0.45 \%$ for $X$ errors. Both of them are close to the standard surface-code threshold (approximately $0.67 \%$) with the added benefit of being suitable for superconducting hardware which significantly reduces issues arising from frequency collisions.  Our schemes are optimized for architectures using the CR gates, but are also similarly useful for other architectures such as those using the controlled-phase gates. Note that for the heavy hexagon code, since $Z$ errors are corrected using a Bacon-Shor type decoding scheme, there is no threshold for such errors. However, low logical errors were observed for the code distances that were considered ($d \le 13$).

More generally, our work here extends the previous fault-tolerant quantum computation schemes with flag qubits, which were mainly in the context of small-size codes, to the realm of topological and subsystem codes. The decoding scheme that we introduce is scalable and can be efficiently implemented for large code distances. We have also proved that there exists topological stabilizer codes with flag qubits defined on a general genus-$g$ surface with gapped boundaries and holes, such that our flag decoder can achieve fault tolerance up to the full code distance.

The paper is organized as follows. In \cref{subsec:heavyhex,subsec:heavysquare} we give a complete description of the heavy hexagon and heavy square codes by describing the gauge generators to be measured and their construction. In addition, we describe the two dimensional layout and decoding graphs of both code families and provide a scheduling for the CNOT gates which minimizes the circuit depths for the $X$ and $Z$-type parity measurements. A more detailed analysis of how edge weights for the Bacon-Shor and Surface-code type decoding graphs are calculated is provided in \cref{app:EdgeWeightCalc}. In \cref{subsec:crossres} we discuss the implementation of the heavy hexagon and heavy square codes using the cross resonance gate and discuss how frequencies can be assigned to different qubits to increase the yield during the fabrication process. Numerics comparing the average number of frequency collisions for the heavy hexagon, heavy square and rotated surface code are provided. In \cref{sec:FlagSection} we provide a detailed description of the decoding algorithm for topological codes which makes use of information from flag qubit measurement outcomes to correct errors. We also discuss how the decoder can be applied to topological codes on a high-genus surface and topological codes with hole defects (more details are provided in \cref{subsec:HighGenus,subsec:HoleDefectsApp}). In \cref{sec:Numerics}, we provide numerical results for the logical failure rates of the heavy hexagon and heavy square codes and provide an estimate of their threshold values. Lastly, in \cref{sec:conclusion} we summarize our results and provide directions for future work.

\section{Heavy hexagon and heavy square codes}
\label{sec:HeavyHex}

\begin{figure}[H]
	\centering
	\includegraphics[width=0.4\textwidth]{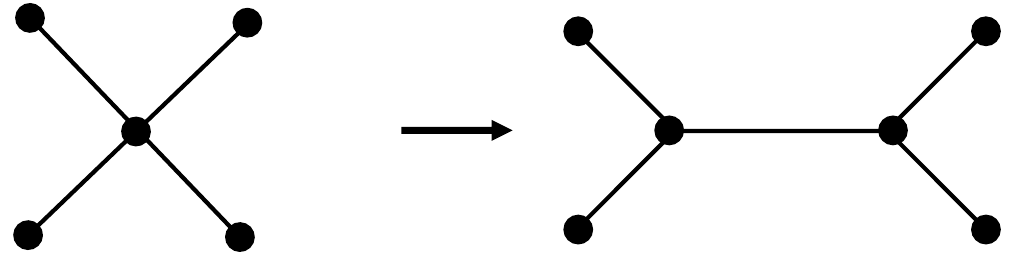}
	\caption{Reduction of a degree four vertex to two vertices of degree three. The vertices represent ancillas and data qubits of some topological code. By adding an additional ancilla qubit and entangling with the original ancilla, the degree of the connectivity can be reduced by one.}
	\label{fig:DegreeReduction}
\end{figure}

Suppose that we have a family of topological codes where the qubits and ancillas are represented as vertices of some graph and the edges of the graph represent the connectivity between the qubits and ancillas. Given a vertex of degree four, it is always possible to reduce the degree to three by adding additional ancilla qubits as shown in \cref{fig:DegreeReduction}. By reducing the degree of the connectivity of a given graph, we will show below that this can potentially reduce the number of frequency collisions that can occur when applying two qubit gates using a cross-resonance interaction \footnote{The larger the degree of a vertex, the more frequencies will be required to apply two qubit gates using the cross-resonance interaction.}.

\begin{figure*}
	\centering
	\includegraphics[width=1.05\textwidth]{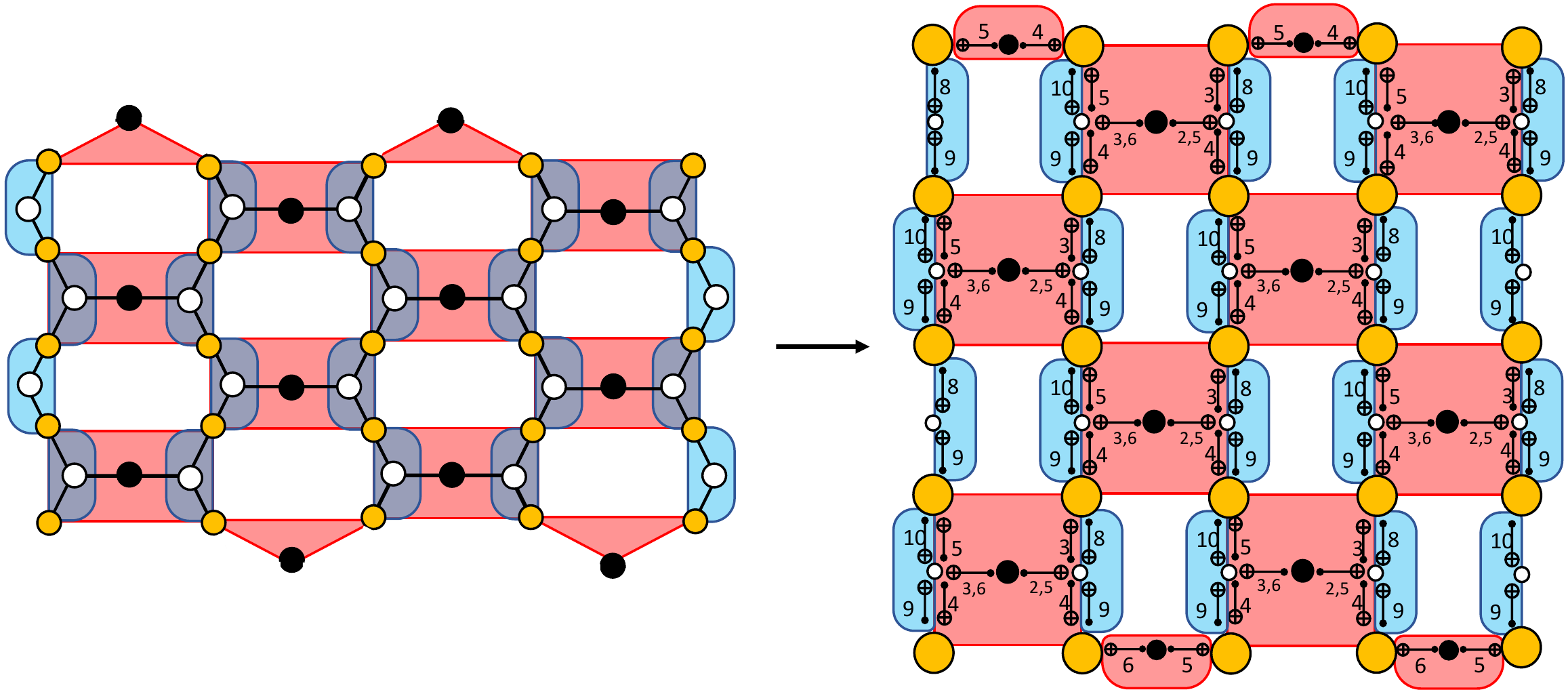}
	\caption{The left of the figure corresponds to the actual layout of the $d=5$ heavy hexagon code which encodes one logical qubit. The data qubits are represented by yellow vertices, white vertices are the flag qubits and dark vertices represent the ancilla to measure the $X$-type gauge generators (red areas) and the $Z$-type gauge generators (blue areas). In the bulk, products of the two $Z$-type gauge generators at each white face forms a $Z$-type stabilizer. The right of the figure provides a circuit illustration of the heavy hexagon code with the scheduling of the CNOT gates used the measure the $X$-type and $Z$-type gauge generators.}
	\label{fig:HeavyHexCircuit}
\end{figure*}

\begin{figure*}
	\centering
	\includegraphics[width=1\textwidth]{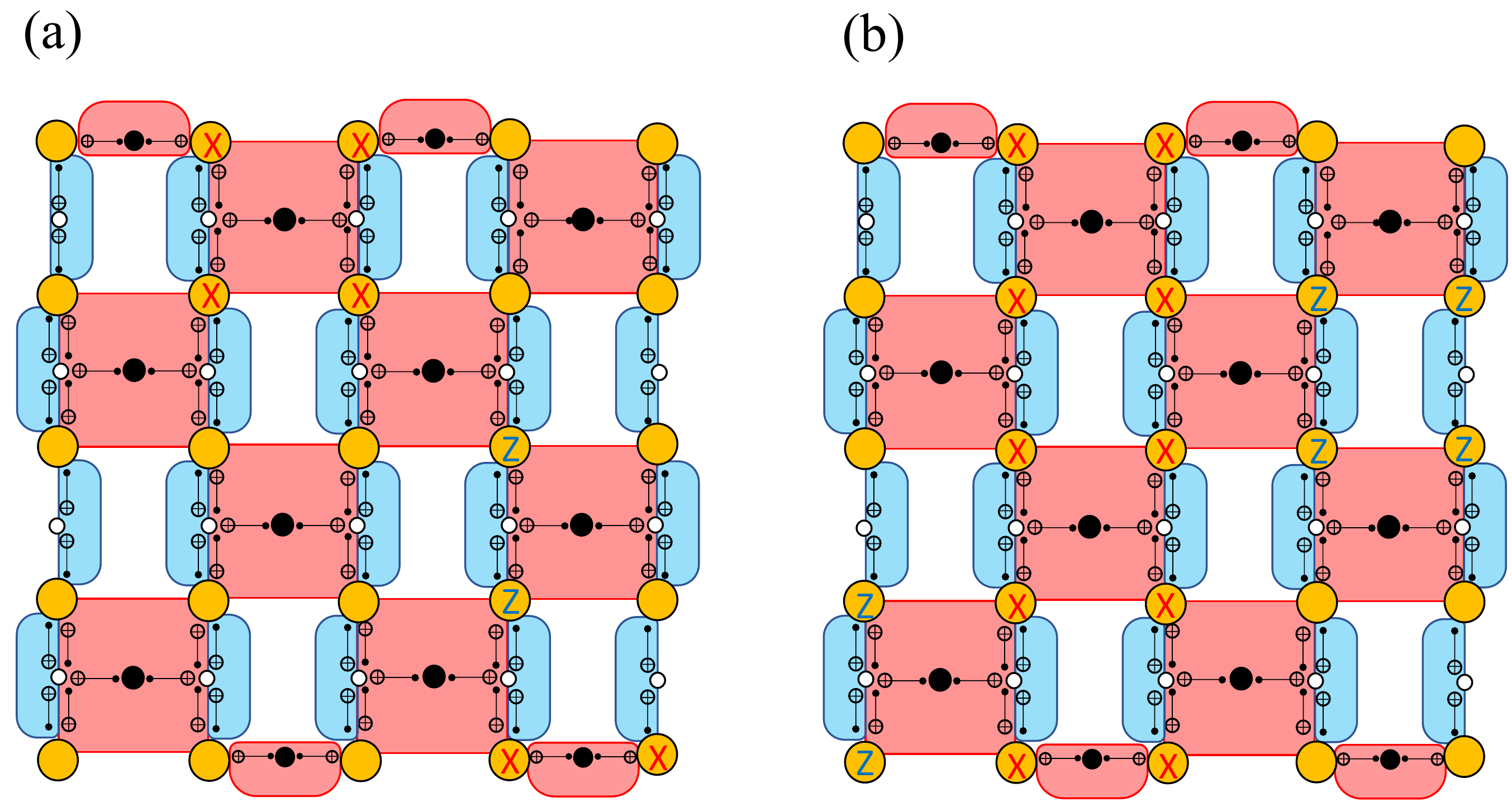}
	\caption{(a) Gauge generators: weight-four X-type in the bulk, weight-two X-type on the upper and lower boundaries, and weight-two Z-type. (b) Stabilizer operators:  a two-column vertical strip with X-type, weight-four Z-type in the bulk, and weight-two Z-type on the left and right boundaries.}
	\label{fig:stabilizers}
\end{figure*}

\subsection{Heavy hexagon code}
\label{subsec:heavyhex}

In this subsection we describe a code family, the heavy hexagon code, encoding one logical qubit and defined on a heavy hexagonal lattice. The adjective ``heavy" is used to say that qubits are placed on both the vertices and edges of a hexagonal lattice. About $60 \%$ of the qubits are therefore degree-2 (i.e.~can interact with just two other qubits), while the rest are degree-3. The average qubit degree is then just $12/5$, a large improvement over the degree-4 square lattice traditionally used for the topological code standard, the surface code. This is also an improvement over the degree-3 connectivity required to implement the Bacon-Shor code on a hexagonal lattice (see e.g. the Bacon-Shor layout in Appendix C of \cite{Yoder2017surfacecodetwist}). An illustration of the distance five heavy hexagon code, along with the scheduling of the CNOT gates for syndrome extraction, is shown in \cref{fig:HeavyHexCircuit}. The data qubits (yellow vertices) in this code, which are used for the encoding of the logical information, reside on an effective square lattice. Hence, the data qubits can be labeled by row and column indices $(i,j)$.

The heavy hexagon code is a subsystem stabilizer code \cite{PoulinSubsystem05,Bacon:2006ul, Aliferis:2007dx}.  In this case, the logical information is encoded and protected in a subsystem with Hilbert space $\mathcal{H_L}$ lying inside a larger Hilbert space,
$\mathcal{H}$$=$$(\mathcal{H_L}\otimes \mathcal{H_G}) \oplus \mathcal{H}_{R}$, where $\mathcal{H_G}$ describes the additional gauge subsystem not necessarily protected against noise and $\mathcal{H}_{R}$ the rest of the full Hilbert space.

The gauge group of the heavy hexagon code is
\begin{align}
\nonumber \mathcal{G}=& \langle Z_{i,j}Z_{i+1,j}, \ X_{i,j}X_{i,j+1}X_{i+1,j}X_{i+1,j+1}, \\
& \quad  X_{1,2m-1}X_{1,2m},  \ X_{d,2m}X_{d,2m+1} \rangle
\end{align}
(with $i,j=1,2,\cdots, d-1$, $m=1, 2, \cdots, \frac{d-1}{2}$,  and the constraint that $i+j$ is even for the second term), which is generated by weight-two Z-type gauge generators (blue areas), weight-four X-type gauge generators (red areas) in the bulk, and weight-two X-type gauge generators (red areas) on the upper and lower boundaries, as illustrated in \cref{fig:stabilizers}(a). Here, $d$ is the code distance, and is taken to be odd throughout the paper in order to optimize the logical error rate. The Z-type and X-type gauge generators are used to correct bit-flip and phase errors respectively.

The stabilizer group which specifies the logical subspace $\mathcal{H_L}$ is the center of the gauge group or, explicitly,
\begin{align}
\nonumber \mathcal{S}=&\langle  Z_{i,j}Z_{i,j+1}Z_{i+1,j}Z_{i+1,j+1}, \
 Z_{2m, d}Z_{2m+1, d}, \\
 &Z_{2m-1, 1}Z_{2m, 1}, \ \prod_i X_{i,j}X_{i,j+1}  \rangle
\end{align}
(with the constraint $i+j$ is odd for the first term).  Here, $Z_{i,j}Z_{i,j+1}Z_{i+1,j}Z_{i+1,j+1}$ is a weight-four surface-code type stabilizer in the bulk, which can be measured via taking the product of the measured eigenvalues of the two weight-two gauge generators $Z_{i,j}Z_{i+1,j}$ and $Z_{i,j+1}Z_{i+1,j+1}$. As will be seen below, the way $Z$-stabilizers can be factorized greatly reduces the circuit depth for syndrome measurements, and hence significantly suppresses the error propagation. In addition to the bulk stabilizers, weight-two surface-code type stabilizers lie on the left and right boundaries. On the other hand, $\prod_i X_{i,j}X_{i,j+1}$ is a Bacon-Shor type stabilizer \cite{Bacon:2006ul, Aliferis:2007dx}, where the Pauli-X operators are supported on a two-column vertical strip, as illustrated in \cref{fig:stabilizers}(b). It can be measured via taking the product of the measured eigenvalues of all the weight-four bulk X-type gauge generators and weight-two boundary X-type gauge generators lying inside the strip.  All the operators inside the gauge group $\mathcal{G}$ commute with the stabilizers in the group $\mathcal{S}$, which are themselves mutually commuting. However, the overlapping gauge operators with different Pauli types do not necessarily commute. Therefore, only the stabilizer eigenvalues are used to infer the errors.  The heavy hexagon code can be considered as a hybrid surface/Bacon-Shor code, where the X and Z errors can be corrected respectively with the surface-code type and Bacon-Shor type decoding procedure respectively. The surface-code part corresponding to the X-error correction is a classical topological code\footnote{A classical topological code stores a classical logical bit string $\overline{Z}$ supported on a non-trivial homological equivalence class of strings, e.g.~connecting two boundaries or form a non-contractible loop around a genus. In particular, the logical string can be deformed to a homologically equivalent one by multiplying the Z-stabilizers. Ref.~[25] has the definition of a classical toric code as an example, which is also the relevant one for the X-error correction of the heavy hexagon code. }, and we will show that the flag qubit measurement outcomes can be used to ensure that the code can correct errors up to the full code distance (see \cref{sec:FlagSection}).

We note that in \cite{CompassCodes}, compass codes, which are defined as gauge-fixes of Bacon-Shor codes, were studied for the purpose of dealing with asymmetric noise models. Such codes include rotated surface codes and Bacon-Shor codes. Thus ignoring the extra ancilla qubits of the heavy hexagonal lattice, the heavy hexagon code can be viewed as belonging to the compass code family.

In general, a distance $d$ version of the code will have $d$ data qubits along each row and each column of the hexagonal lattice so that the code parameters are given by $\codepar{d^2,1,d}$. In addition, a distance $d$ implementation of the code requires a total of $\frac{d+1}{2}(d-1)$ syndrome measurement qubits and $d(d-1)$ flag qubits. Hence the total number of qubits in the implementation of the code is $\frac{5d^2-2d-1}{2}$.

\begin{figure}
	\centering
	\includegraphics[width=0.5\textwidth]{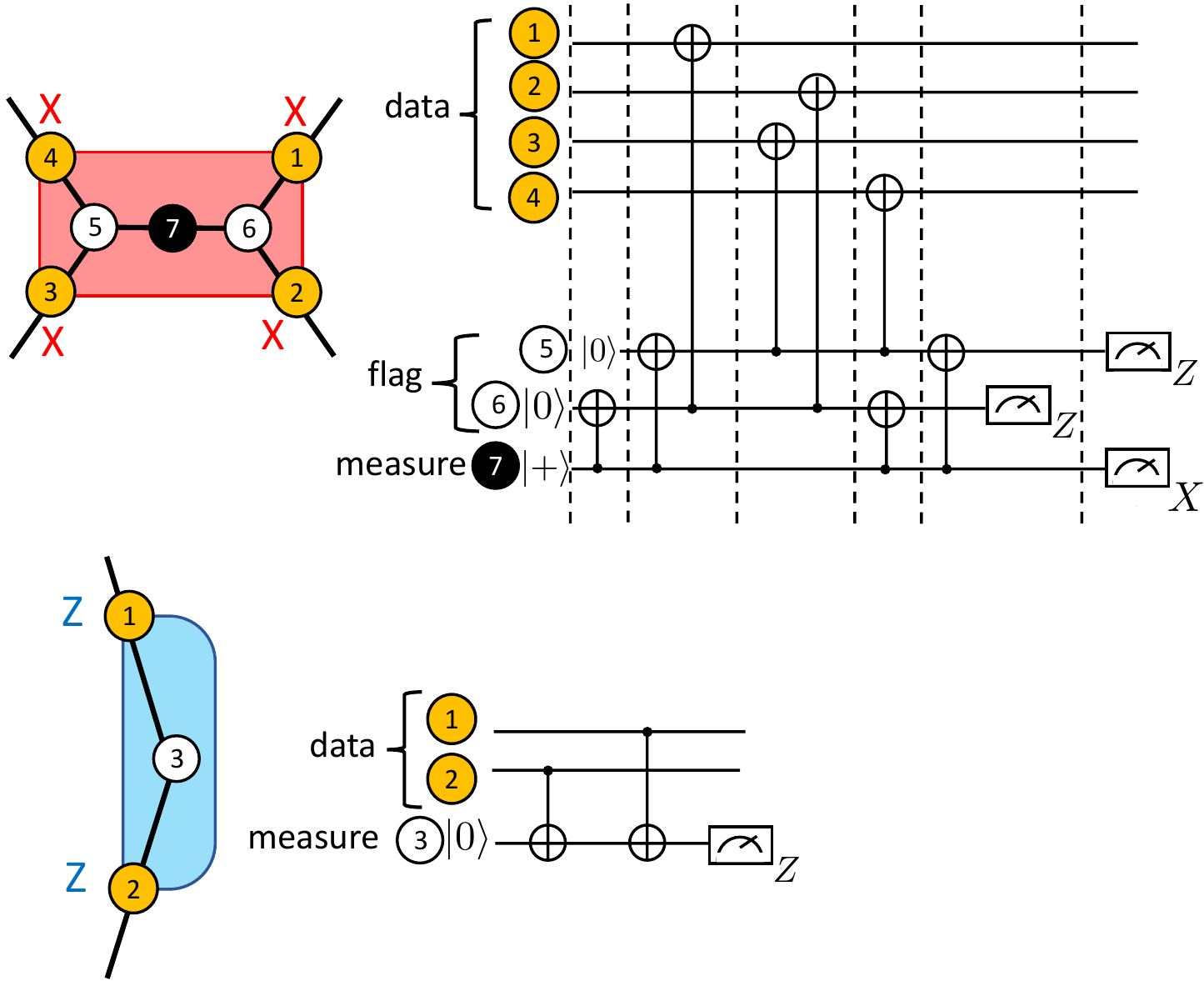}
	\caption{Circuit to perform the $X$ and $Z$-type parity measurements of the heavy hexagon code. Two flag qubits (white circles) are used to measure the weight-four $X$-type gauge generators.}
	\label{fig:HeavyHexStabMeasurements}
\end{figure}

Complementing the righthand side of \cref{fig:HeavyHexCircuit}, an illustration of the circuits for measuring the $X$ and $Z$-type gauge generators is given in \cref{fig:HeavyHexStabMeasurements}.
The CNOT scheduling was chosen to minimize the total number of error locations for one round of syndrome measurements. Such a scheduling is implemented in 11 time steps, which includes qubit initialization and measurement. 

\begin{figure*}
	\centering
	\subfloat[\label{fig:BaconGraph}]{%
		\includegraphics[width=0.27\textwidth]{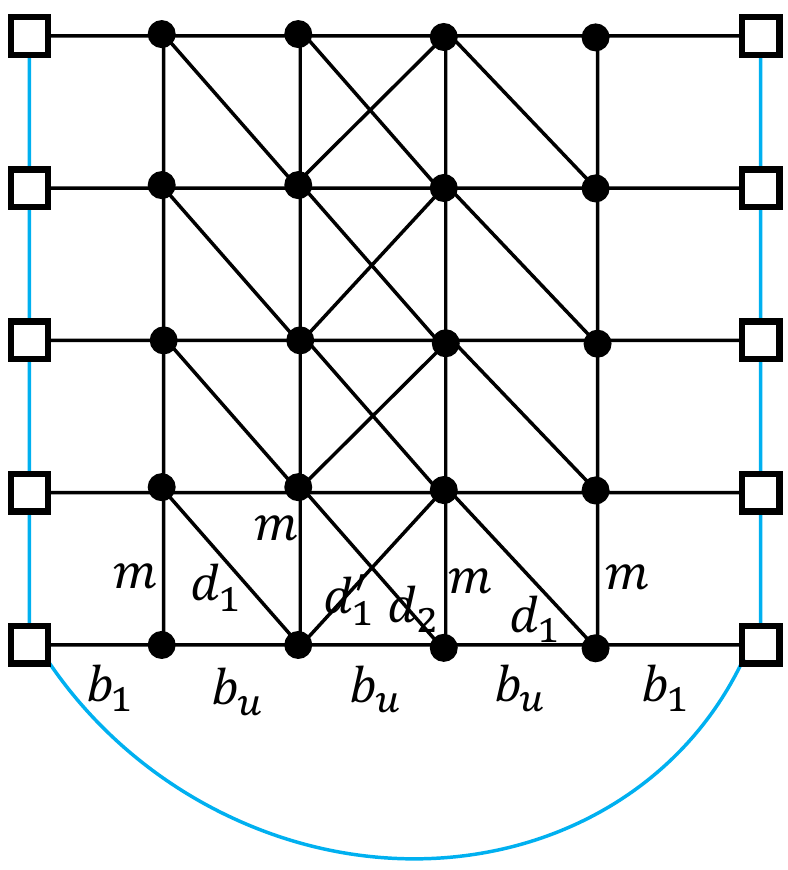}
	}
	\subfloat[\label{fig:ZHexGraph}]{%
		\includegraphics[width=0.38\textwidth]{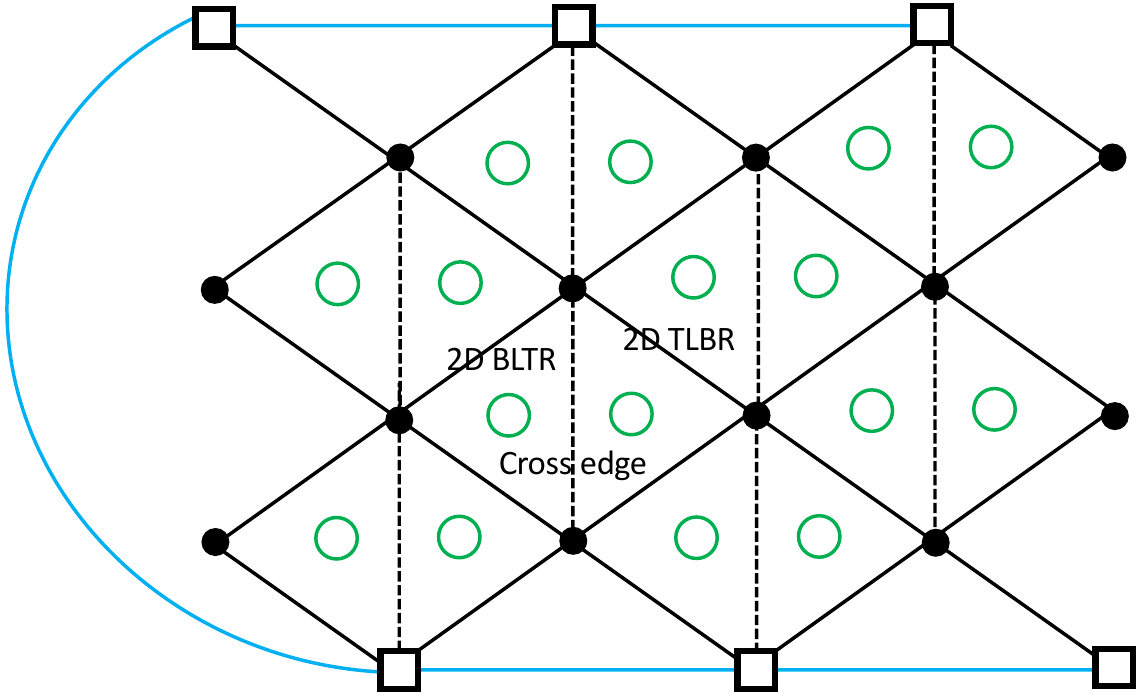}
        }
	\subfloat[\label{fig:ZHexGraphWithDiag}]{%
		\includegraphics[width=0.38\textwidth]{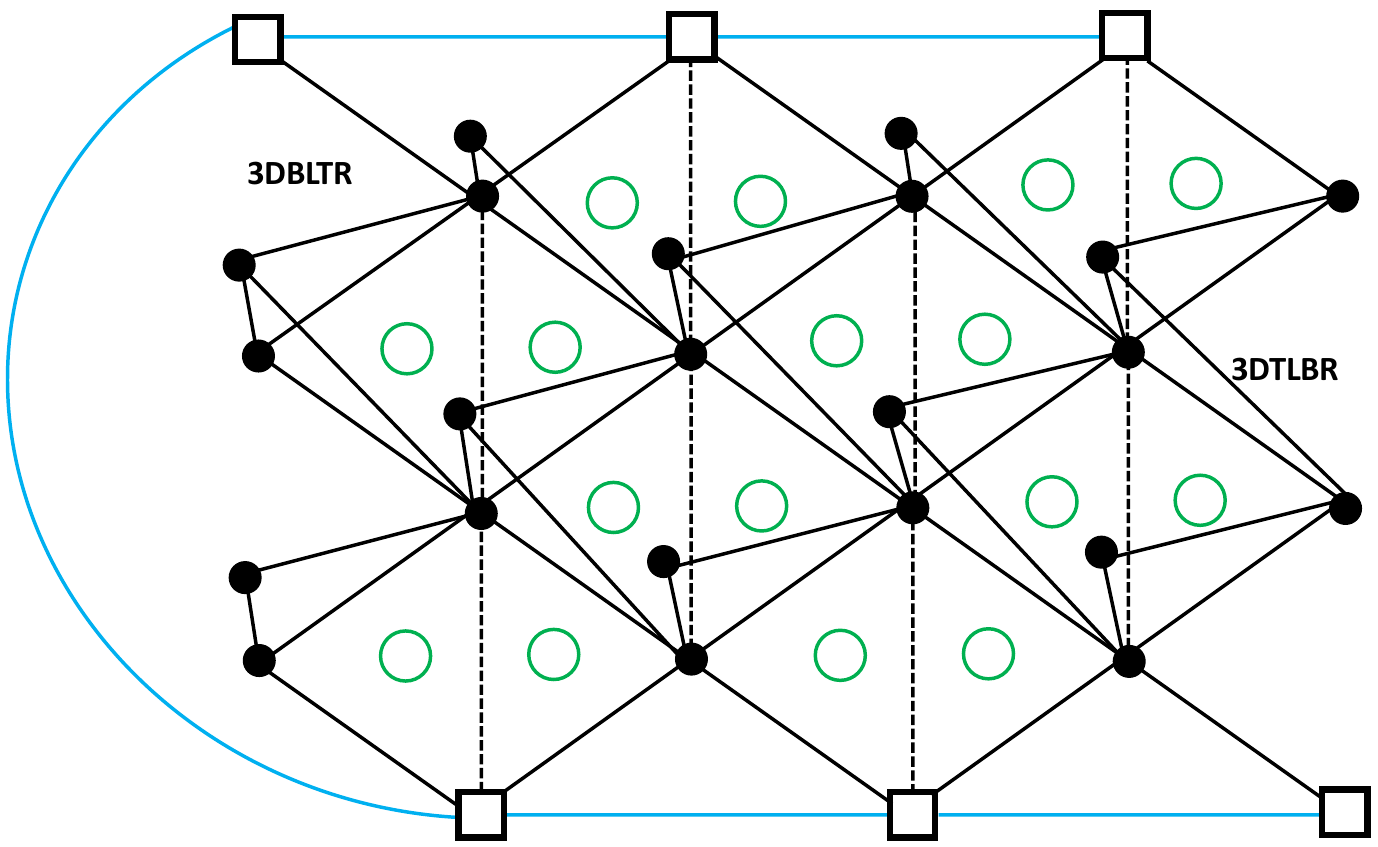}
	}
	\caption{(a) Example of the minimum weight matching graph for five rounds the X-type gauge measurements of the $d=5$ heavy hexagon code. Since only the stabilizer measurements are used to correct errors, the graph is one-dimensional for one measurement round. Diagonal edges connecting two consecutive one-dimensional graphs are necessary to correct space-time correlated errors arising from CNOT gate failures. The weights and directions of the diagonal edges depends on the CNOT gate scheduling and are chosen such that a single fault in the measurement circuits corresponds to an edge in the final graph. Labels $m$, $b_1$, $b_u$, $d_1$, $d'_1$ and $d_2$ correspond to edges with different weights computed based on the probabilities of occurrence for a given edge. The edge $d'_1$ is a bulk feature that only appears in odd columns excluding the first and last column (for instance, columns three and five in the $d=7$ graph would contain edges $d'_1$). More details are given in \cref{app:EdgeWeightCalc}. (b) Example of the minimum weight matching graph for one round of the Z-type stabilizer measurements of the $d=5$ heavy hexagon code. The full graph for $d$ rounds of $Z$-type stabilizer measurements is three-dimensional. Cross edges are given by dashed lines since they are only present in the presence of non-trivial flag measurement outcomes (during the $X$-type gauge measurements) represented by green circles. More details are provided in \cref{sec:FlagSection}. (c) Diagonal edges connecting two-dimensional graphs are added to ensure that a single fault in the measurement circuits corresponds to an edge in the final graph.}
	\label{fig:HeavyHexGraphs}
\end{figure*}

Although the circuit depth of the heavy hexagon code is larger than that of rotated surface code (which requires a total of six time steps for the $X$ and $Z$ stabilizer measurements) \cite{FMMC12,TS14}, the flag qubits can be used to correct weight-two errors arising from a single fault during the weight-four $X$-type gauge measurements. In \cref{sec:FlagSection}, we provide a new decoding algorithm which uses information from the flag measurement outcomes to correct errors up to the full distance of the code\footnote{For a distance $d$ code, the algorithm will allow any error arising from at most $\floor*{\frac{d-1}{2}}$ faults to be corrected.}.

An important ingredient in the implementation of minimum weight perfect matching using Edmond's method \cite{Edmonds65} is the matching graphs used to correct $X$ and $Z$-type Pauli errors with appropriate edges and edge weights. We illustrate such two-dimensional graphs for the distance-five heavy hexagon code in \cref{fig:HeavyHexGraphs}. The graphs are constructed by assigning an edge to each data qubit and vertices for the ancilla qubits used to measure the $X$-type gauge generators (\cref{fig:BaconGraph}) and $Z$-type stabilizer generators (\cref{fig:ZHexGraph}). The blue edges are boundary edges which have zero weight. In general, edge weights are given by $w_{\text{E}} = -\log{P_{\text{E}}}$ where $P_{\text{E}}$ is the total probability of error configurations resulting in an error at the edge $\text{E}$ (see \cite{FowlerEdgeWeights} for examples of optimizations performed on the surface code and \cref{app:EdgeWeightCalc} for edge weight calculations of the heavy hexagon and heavy square code). Correctable $Z$-type errors will result in highlighted vertices of the graph in \cref{fig:BaconGraph} while correctable $X$-type errors will result in highlighted vertices of the graph in \cref{fig:ZHexGraph}.  If an odd number of vertices are highlighted, a square vertex (chosen at random) is highlighted to ensure that the total number of highlighted vertices is always even. Note that for one round of syndrome measurements, the matching graph for $X$-type stabilizers is one dimensional since a vertex along a column is highlighted if the sum of the measured eigenvalues of all $X$-type gauge generators along a strip is odd.To detect measurement errors, the gauge measurements must be repeated $d$ times \cite{DKLP02,FowlerMatching,FMMC12}. Thus the matching graph consists of $d$ copies of the matching graph for one round of gauge measurements with vertices connected by vertical edges. For the $X$-type gauge measurements, we obtain a two-dimensional graph whereas for the $Z$-type stabilizers, we obtain a three-dimensional graph. In addition, diagonal edges connecting the graphs from two consecutive measurement rounds must be added to ensure that any single fault in the circuits of \cref{fig:HeavyHexStabMeasurements} corresponds to an edge in the final graph. Minimum wight perfect matching is then applied on the subgraph of highlighted vertices.

\subsection{Heavy square code}
\label{subsec:heavysquare}

\begin{figure}
	\centering
	\includegraphics[width=0.5\textwidth]{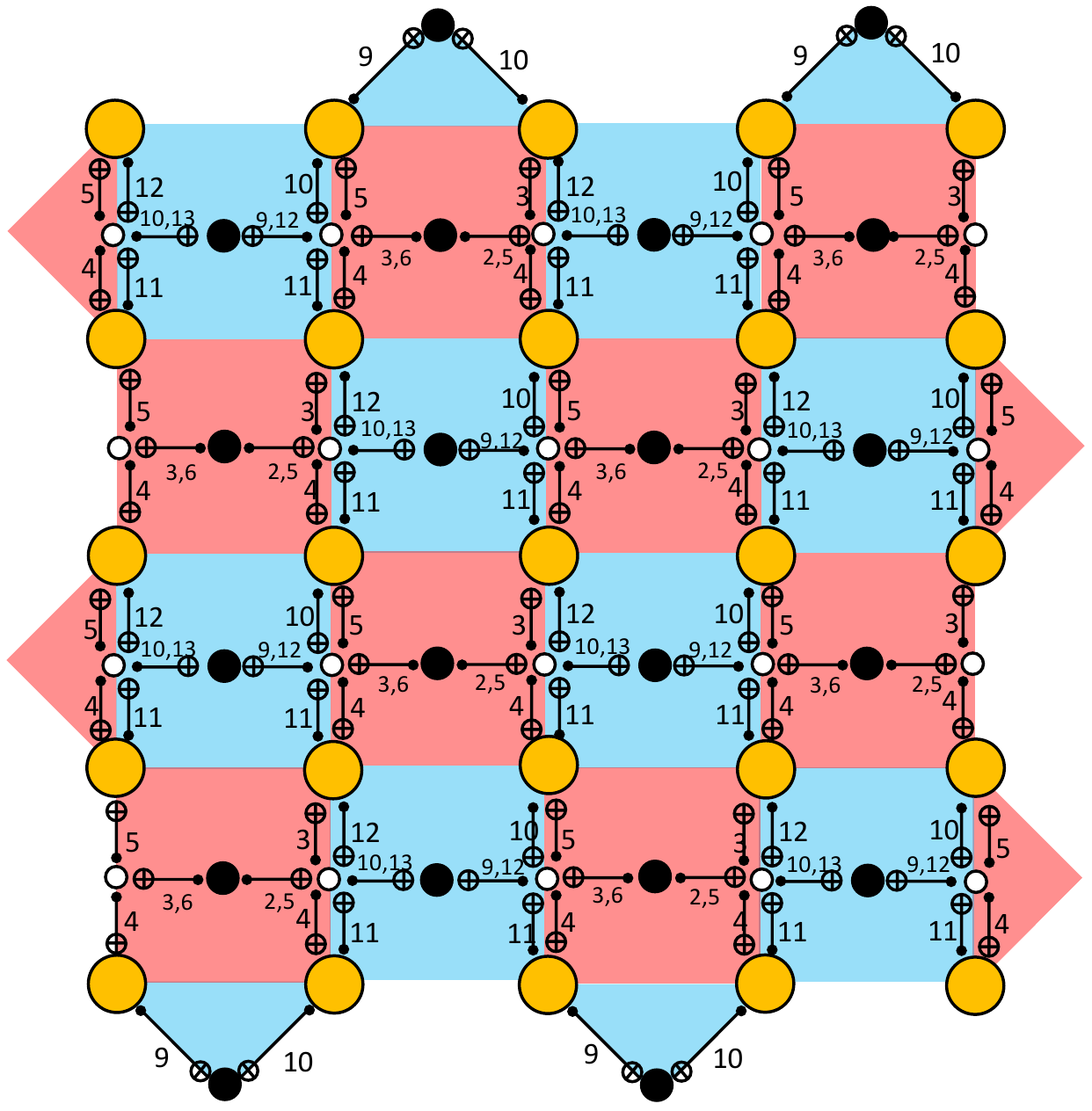}
	\caption{Illustration of the $d=5$ heavy square code with the scheduling of the CNOT gates. The data qubits are represented by yellow vertices, white vertices are the flag qubits and the dark vertices are the syndrome measurement qubits. The red faces correspond to $X$-stabilizer measurements and the blue faces to $Z$-stabilizer measurements. }
	\label{fig:HeavySquareLattice}
\end{figure}

\begin{figure}
	\centering
	\includegraphics[width=0.5\textwidth]{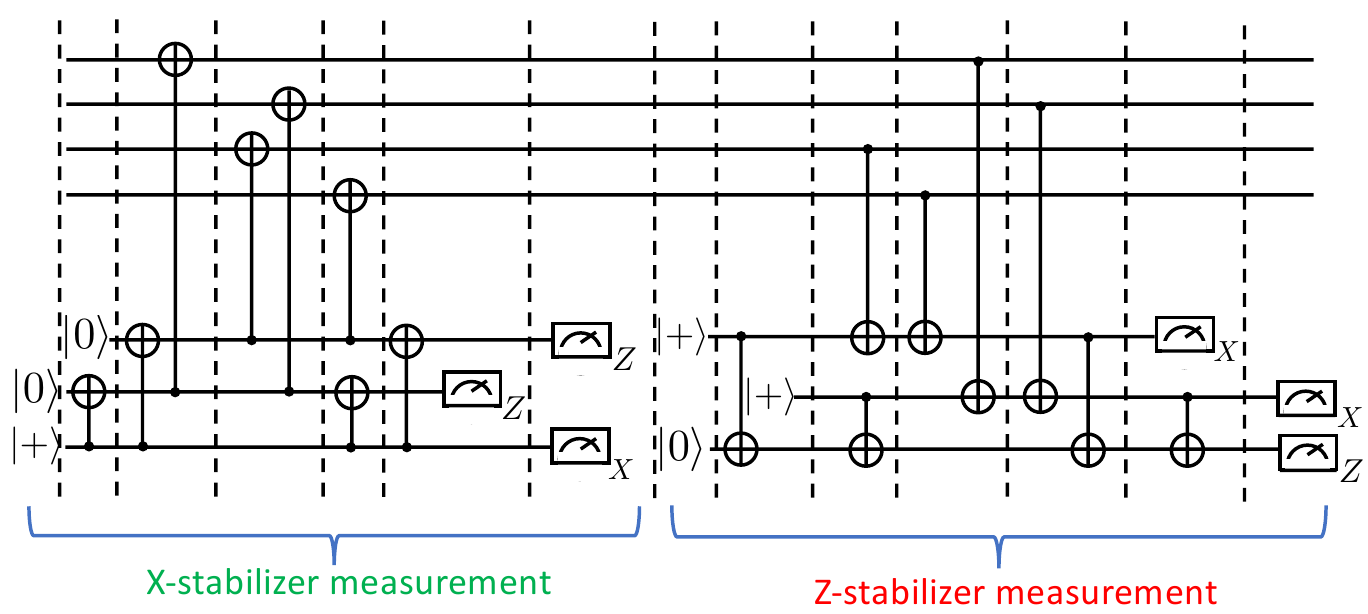}
	\caption{Circuit for the measurement of the $X$-stabilizers followed by the $Z$-stabilizers of the heavy square code. }
	\label{fig:HeavySquareXZStab}
\end{figure}

\begin{figure*}
		\centering
	\subfloat[\label{fig:ZStabHeavySquareMatchGraph}]{%
		\includegraphics[width=0.31\textwidth]{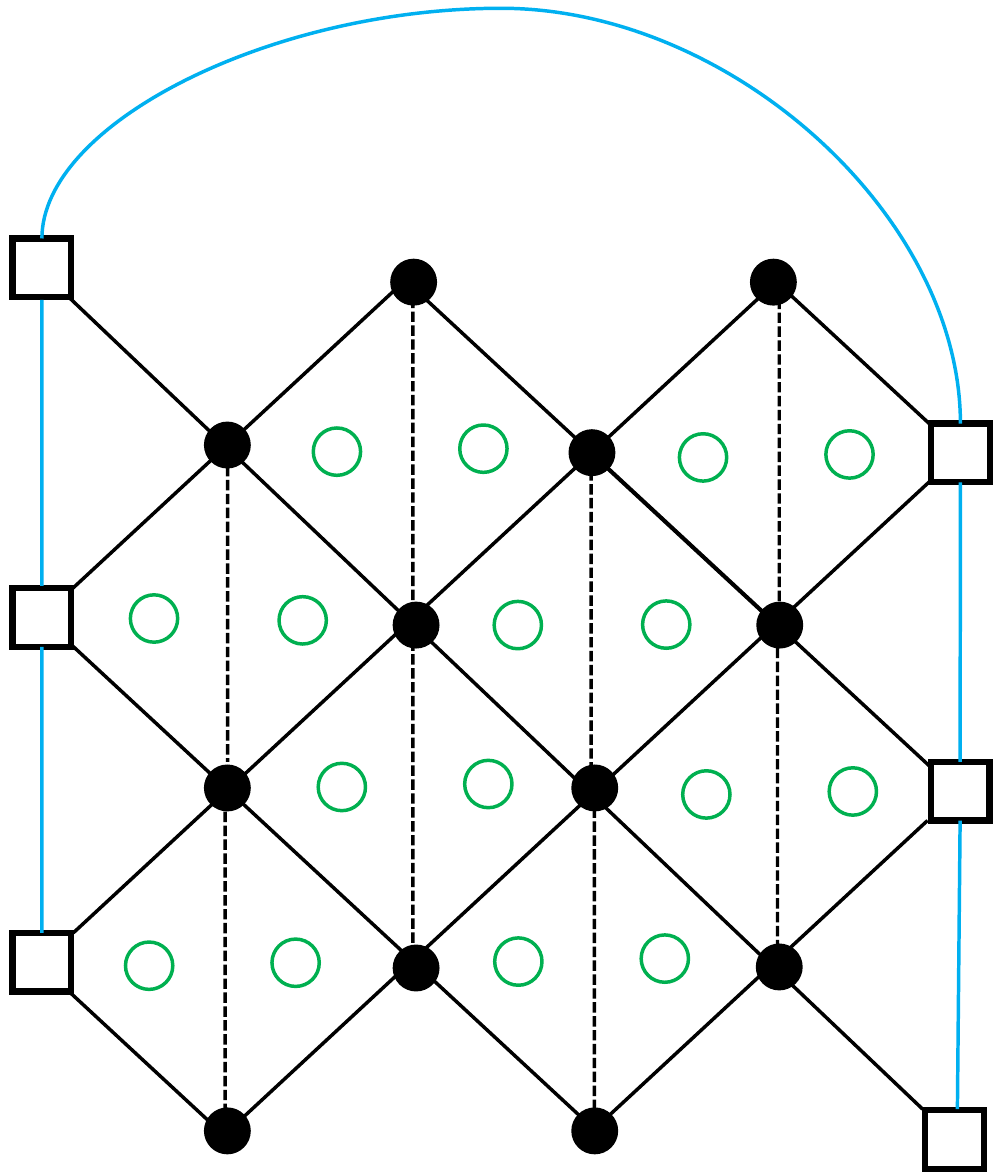}
	}
	\subfloat[\label{fig:XStabHeavySquareMatchGraph}]{%
		\includegraphics[width=0.34\textwidth]{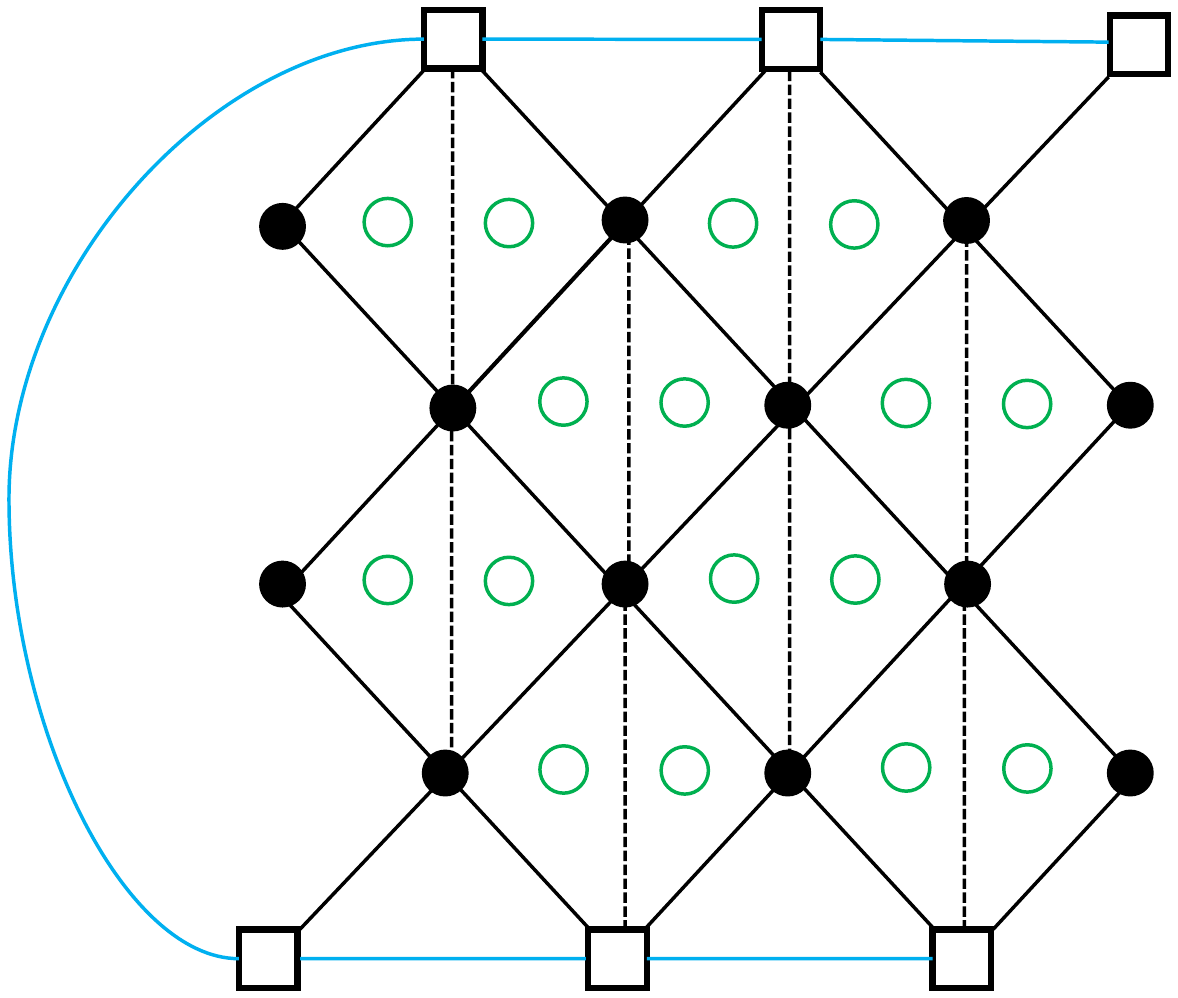}
        }
	\subfloat[\label{fig:HeavySuqare3DdiagGraph}]{%
		\includegraphics[width=0.34\textwidth]{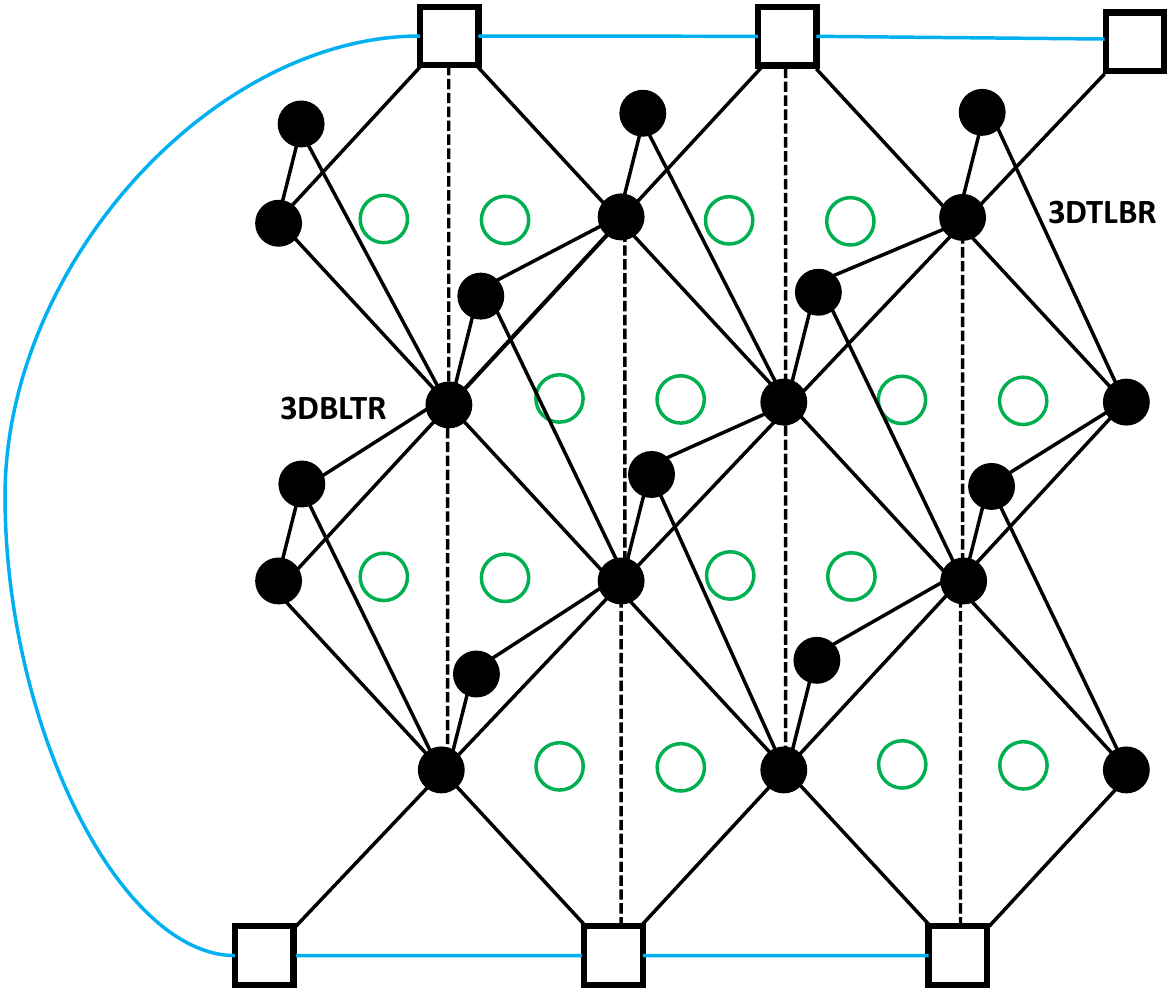}
	}		
	\caption{Example of the graphs used to implement minimum weight perfect matching for the $d=5$ heavy square code for (a) $Z$-type stabilizer measurements and (b) $X$-type stabilizer measurements. The green vertices in (a) correspond to flag measurement outcomes during $X$-stabilizer measurements. Similarly, the green vertices in (b) correspond to flag measurement outcomes during $Z$-stabilizer measurements. In (c) we illustrate the graph associated with $Z$-stabilizer measurements with 3D diagonal and vertical edges connecting the two-dimensional graphs.}
	\label{fig:MatchGraphHeavySquare}
\end{figure*}

In this subsection, we present a mapping of the surface code onto the heavy square lattice. In the implementation of the rotated surface code in \cite{TS14}, each qubit in the bulk, both data and syndrome measurement qubits, interacts with four qubits. By adding two flag qubits to each stabilizer measurement in the bulk, the rotated surface code can be mapped onto a heavy square lattice as shown in \cref{fig:HeavySquareLattice} where now the average qubit degree is $8/3$. The distance $d$ code belonging to the family has parameters $\codepar{d^2,1,d}$ with $d^2$ data qubits and $2d(d-1)$ flag and syndrome measurement qubits. Hence the total number of qubits required for the implementation of the code is $3d^2 - 2d$. In what follows, the code family will be referred to as the heavy square code. With the addition of the flag qubits, it can be seen that the degree of both ancilla and data qubits has been reduced to two in the bulk, at the cost of having flag qubits with degree four. In \cref{subsec:crossres}, more details will be provided showing that the heavy square code does reduce the number of frequency collisions relative to the standard implementation of the surface code.

\begin{figure}
	\centering
	\includegraphics[width=0.51\textwidth]{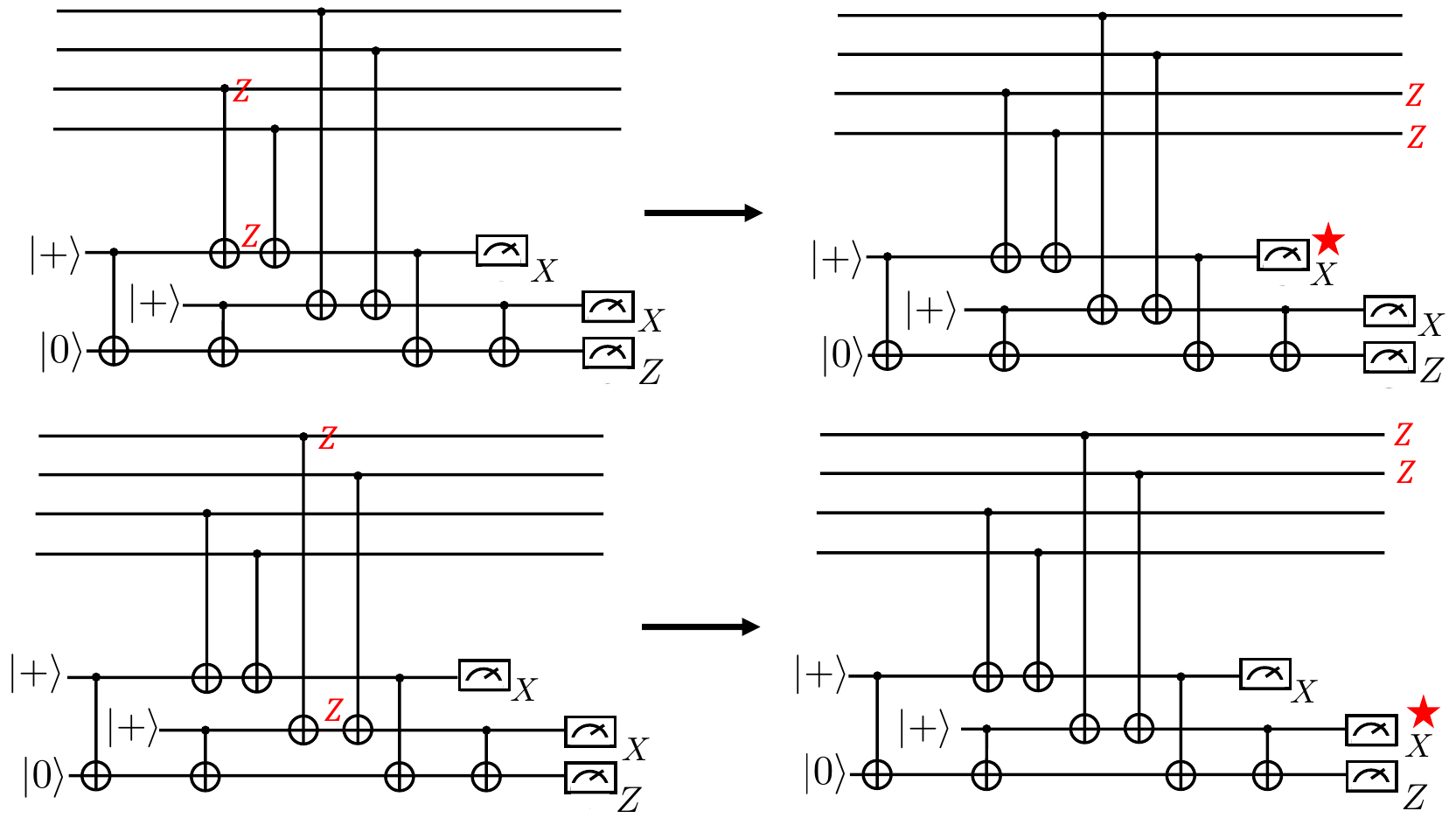}
	\caption{Flag outcomes arising from a single fault resulting in a weight-two data qubit error during the measurement of a weight-four operator. In this example the flag qubits are prepared in the $\ket{+}$ state and measured in the $X$ basis. Starred measurements give a nontrivial outcome due to the errors.}
	\label{fig:FlagOutcomes}
\end{figure}

\begin{figure*}
	\centering
	\includegraphics[width=0.8\textwidth]{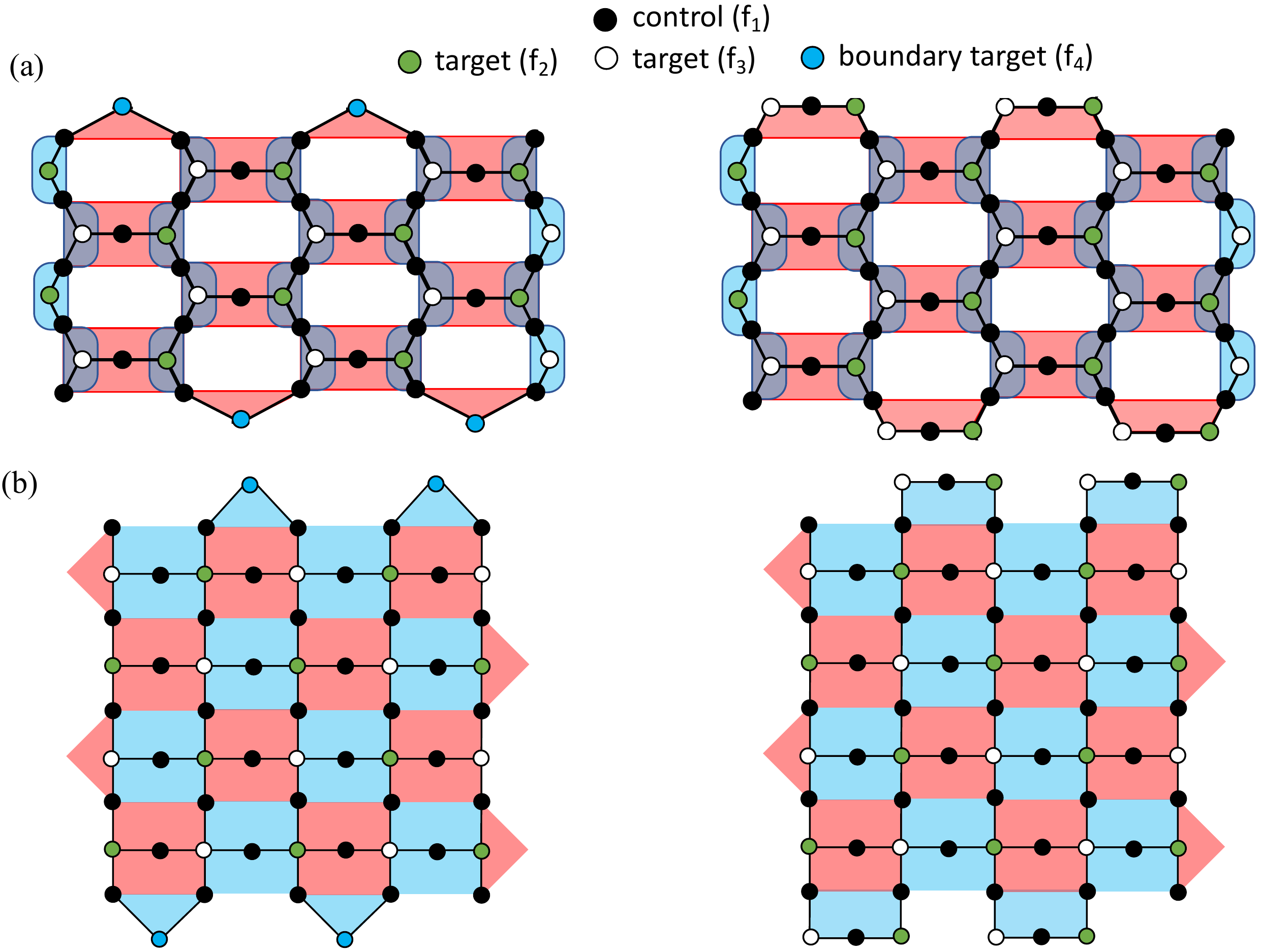}
	\caption{Frequency assignments of the heavy hexagon code (a) and the heavy square code (b). Solid black lines indicate the actual connections and CNOT gates in the fabricated superconducting device. The black dots indicate control qubits,  while the other three colors indicate target qubits. The original heavy hexagon and heavy square codes on the left panels feature three distinct frequencies in the bulk and an additional frequency on the boundary, while the modified codes on the right panels remove the extra frequencies (white dots) on the boundaries.}
	\label{fig:frequency_assignment}
\end{figure*}

The circuit for measuring the $X$ and $Z$ stabilizers of the heavy square code in the bulk, following the CNOT scheduling of \cref{fig:HeavySquareLattice}, is illustrated in \cref{fig:HeavySquareXZStab}. It can be seen that the total number of time steps required to perform the stabilizer measurements is 14, compared to 11 for the heavy hexagon code. The increase in the number of time steps compared to the heavy hexagon code is due to the fact that both $X$ and $Z$ stabilizers have weight four compared to the weight-two $Z$-type gauge generators of the heavy hexagon code. Examples of the matching graphs for the $d=5$ heavy square code is given in \cref{fig:MatchGraphHeavySquare}. An illustration of the possible weight-two errors arising from a single fault along with the flag qubit outcomes is given in \cref{fig:FlagOutcomes}.

We conclude this section with an important remark regarding the role of the flag qubits for the heavy square code. The logical $\overline{X}$ operator of the heavy square code has minimum support on $d$ qubits along each column of the lattice. From the CNOT scheduling of \cref{fig:HeavySquareLattice}, a weight-two $X$ error arising from a single fault (which will result in a non-trivial flag measurement outcome) will be orthogonal to $\overline{X}$ since its support will remain in one column. On the other hand, a logical $\overline{Z}$ operator of the heavy square code will have support on each row of the lattice. A weight-two $Z$ error arising from a single fault will thus be parallel to $\overline{Z}$ (since just like for $X$ errors, its support will be along a single column). For the surface code, this problem can be avoided by finding a scheduling (see \cite{TS14}) such that weight-two errors arising from a single fault are always orthogonal to the logical operator of the same type. Performing an exhaustive numerical search, such a scheduling is not possible for the heavy square code when the flag qubits are used to reduce the degree of data and syndrome measurement qubits. Hence if the flag measurement outcomes were omitted when decoding the heavy square code, the effective distance of the code $d_{\text{eff}}$ would satisfy $d_{\text{eff}} < d$. In \cref{sec:FlagSection}, we provide a decoder that uses the flag measurement outcomes allowing such weight-two $Z$ errors arising from a single fault (which are parallel to $\overline{Z}$) to be corrected. We will show that with such a decoder, the code can correct any error arising from at most $(d-1)/2$ faults so that $d_{\text{eff}} = d$.

\subsection{The cross resonance gates and frequency collision reduction}
\label{subsec:crossres}

The designs of the heavy hexagon and heavy square codes are motivated by the experimental implementation of fault-tolerant quantum computation with a superconducting architecture.  The low-degree property of the graphs significantly mitigates the issues of frequency collision and crosstalk and is applicable to a wide range of architectures including cross-resonance (CR) gates \cite{Rigetti2010, Chow2011}, controlled phase gates \cite{BKM+14, DiCarlo:2009ja}, and systems with tunable couplers \cite{Chen:2014cw, McKay2016}. In this section, we show that our codes are optimized for a CR gate architecture.

The CR gate implements a CNOT between a control and a target qubit, using only microwave pulses and not the magnetic flux drive needed in other gate types \cite{BKM+14, DiCarlo:2009ja, Chen:2014cw, McKay2016}. When employed to couple fixed-frequency transmon qubits via microwave-resonant buses, this architecture is hardware-efficient and insensitive to the charge-noise-induced dephasing noise source. \cite{Chow2014, Corcoles2015}. The current fidelity of the gate exceeds 0.99 in a 2-qubit setup  \cite{Sheldon2016}, approaching the error threshold for the surface code.  Small-scale multi-qubit demonstrations of fault-tolerant protocols has been achieved recently \cite{Chow2014, Corcoles2015, Takita2016, Takita2017}.

In the CR gate, a drive tone at the target qubit's resonance frequency is applied to the control qubit. This requirement can produce `frequency collisions' among nearby qubits whose energies are degenerate. As transmon qubits are weakly anharmonic, the $\ket{0} \rightarrow \ket{1}$, $\ket{1} \rightarrow \ket{2}$ and $\ket{0} \rightarrow \ket{2}$ transitions are all relevant. Two nearest-neighbor qubits must not have degenerate $\omega_{01}$, nor one qubit's $\omega_{01}$ degenerate with another's $\omega_{12}$ or $\omega_{02}/2$. Among next-nearest-neighbor qubits joined to a common control qubit of the CNOT gate, degeneracies of $\omega_{01}$ and $\omega_{12}$ are also forbidden, as is a control qubit's $\omega_{02}$ being degenerate with the summed $\omega_{01}$ of two of its nearest neighbors. On the other hand, if a control and target's $\omega_{01}$ frequencies are too far apart, the gate rate becomes too slow \cite{MagesanArxiv2019}. To avoid all these collision conditions, we designate each qubit to have one of a minimal set of distinct frequencies according to a defined pattern. The relative frequencies of nearest-neighbor qubits therefore fix the CNOT direction among each pair. In order to reverse certain CNOT directions to implement the measurement circuits in the codes developed above, we can conjugate the existing CNOT by Hadamards on the control and target qubits.  Since the single-qubit errors on current superconducting architectures are at least an order of magnitude lower than the two-qubit gate fidelities, the errors due to these extra Hadamards are negligible.

In \cref{fig:frequency_assignment} we show the qubit frequency assignments of the heavy hexagon and heavy square codes respectively. The solid black lines indicate the connections and CNOT gates on the actual device.  The control qubits are represented by black dots assigned with frequency $f_1$, while the target qubits in the bulk are represented by green and white dots corresponding to frequency $f_2$ and $f_3$.  In addition, in both codes,  there are additional boundary target qubits with frequency $f_4$ represented by blue dots (shown in the left panels).   In both the heavy hexagon and heavy square codes, the controls reside on the degree-two vertices of the graph, such that they only have at most two neighboring targets.   With this configuration, there are only three distinct frequencies ($f_1$, $f_2$ and $f_3$) in the bulk, which greatly reduces the possibility of frequency collisions.   We note that the extra frequency $f_4$ from the boundary targets are due to the modification/simplification of the heavy hexagon and heavy square lattice structure on the boundaries in order to shorten the circuit depth of the boundary gauge or stabilizer generators.  If we recover the original heavy hexagon and heavy square lattice structure on the boundaries at the price of introducing additional ancillas and increase the depth of the measurement circuits (as shown on the right panels of \cref{fig:frequency_assignment}), we will again have only three distinct frequencies (i.e., removing all the blue dots). By contrast, a rotated surface code architecture, in which all qubits reside on degree-four vertices, must have five distinct frequencies in order to avoid all collision conditions. \cite{Gambetta2017}

\begin{figure}
	\centering
	\includegraphics[width=0.5\textwidth]{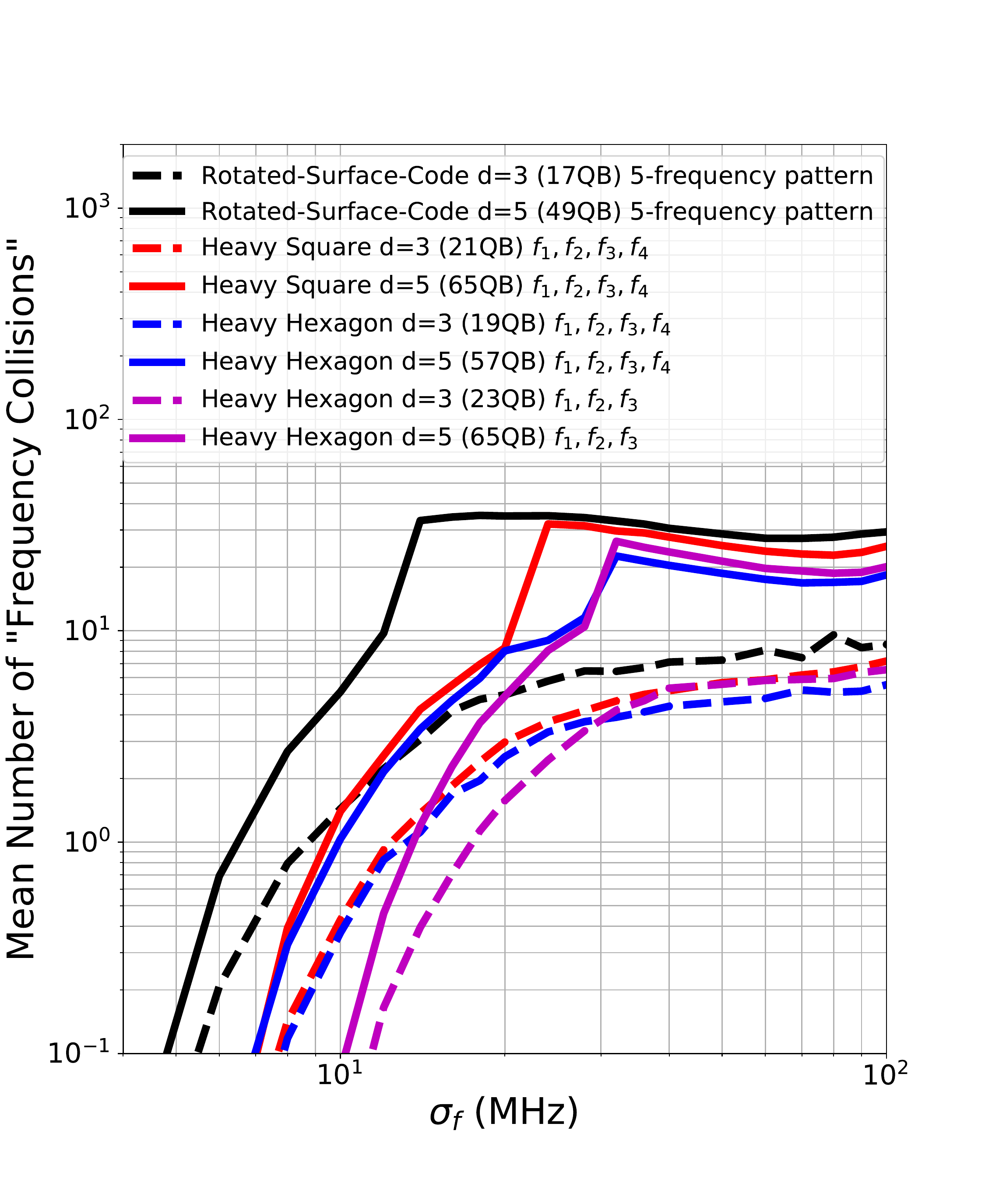}
	\caption{Monte Carlo simulations of the collision rate due to qubit frequency scatter caused by fabrication imprecision. Averages of $>10^3$ Monte Carlo repetitions. The x-axis is the frequency precision of the fabrication, $\sigma_f$ (MHz).  The y-axis is the mean number of collisions for each device for each value of $\sigma_f$.   Heavy hexagon code:  $d=3$ with the 3-frequency design in the right panel of \cref{fig:frequency_assignment}(a)  (magenta dashed line);  $d=5$ with the 3-frequency design in the right panel of \cref{fig:frequency_assignment}(a)  (magenta solid line); $d=5$ with the 4-frequency design in the left panel of \cref{fig:frequency_assignment}(a)  (blue solid line). Heavy square code:  $d=3$ with the 4-frequency design in the right panel of \cref{fig:frequency_assignment}(a)  (red dashed line);  $d=5$ with the 3-frequency design in the right panel of \cref{fig:frequency_assignment}(b)  (red solid line). Rotated surface code, degree 4 and five-frequency pattern: $d=3$ (black dashed line);  $d=5$ (black solid line).}
	\label{fig:frequency_collision}
\end{figure}

For practical implementations, a code with its graph and set of frequencies must be robust against the disorder that develops among dozens of transmon qubits prepared together on a single chip. This disorder, arising from imperfections in fabrication, may be characterized by the parameter $\sigma_f$, the standard deviation in frequency of a population of fixed-frequency transmons. For typical multi-qubit devices, whose transmons have $f_{01} \sim 5$ GHz and $f_{12} - f_{01} \sim -330$ to $-340$ MHz (similar to \cite{Sheldon2016} or \cite{IBMQXYorktown}), achieving $\sigma_f < 50$ MHz requires all device parameters to be controlled with precision better than $1\%$, not a simple task when the transmons incorporate nanoscale tunnel junctions and capacitances of tens of fF. We seek therefore to find a lattice and code for which the transmons may have the largest possible imprecision $\sigma_f$ while still avoiding frequency collisions. For a quantitative comparison among the lattices and frequency patterns of \cref{fig:frequency_assignment}, we perform Monte Carlo simulations in which we populate these lattices and related designs with a random disorder in frequency characterized by $\sigma_f$. Taking `collisions' as described above, we forbid regions of frequency space where we expect the resulting gate errors to exceed other typical causes \cite{Sheldon2016,IBMQXYorktown,MagesanArxiv2019}. In \cref{fig:frequency_collision} we show the mean number of collisions found among various lattices as a function of $\sigma_f$, each case derived from at least $10^3$ Monte Carlo repetitions \cite{HutchingsPhysRevApplied_8_044003}. While this model does not quantify gate errors, it does enable us to compare frequency-crowding in different graphs using consistent means. As a practical matter, we seek to achieve $< 1$ average collision. We see that in any case this goal is achievable only for precisions $\sigma_f < 30$ MHz.

For all three types of codes, i.e., the heavy hexagon code, heavy square code, and the rotated surface code, we plot the mean number of collisions vs $\sigma_f$ for both distance $d=3$ and distance $d=5$.  For the heavy hexagon code, we show the 3-frequency design for both $d=3$ (magenta dashed line) and $d=5$ (magenta solid line) corresponding to the right panel of \cref{fig:frequency_assignment}(a), and also the 4-frequency design at $d=3$ (blue dashed line) and $d=5$ (blue solid line) corresponding to the left panel of \cref{fig:frequency_assignment}(a).  Note that the behavior of the two types of designs at $d=5$ are similar, with the 4-frequency design having slightly more collisions, which is expected since the lattice shape differs only on the boundary.  For the heavy square code, we show the 4-frequency design for both $d=3$ (red dashed line) and $d=5$ (red solid line) corresponding to the right panel of \cref{fig:frequency_assignment}(b). We can conclude that the heavy hexagon and heavy square codes behave similarly for each code distance (red vs blue lines). Each of these is, however, distinctly better than the rotated surface code with degree-four and a five-frequency pattern (black dashed and straight lines). Although the rotated surface code requires 10 to 20\% fewer qubits than the other two codes at each distance $d$, it requires qubits to be prepared nearly twice as precisely in order to eliminate frequency collisions. Or put another way, for a given distance ($d=3$ or $d=5$) and fabrication precision $\sigma_f$, the rotated surface code exhibits roughly an order of magnitude more frequency collisions than do the heavy hexagon or heavy square codes.  Therefore, the design of error correcting codes on a low-degree graph indeed improves the fabrication of the hardware significantly.

\section{Decoding the heavy hexagon and heavy square codes using flag qubits}
\label{sec:FlagSection}

\begin{figure}
	\centering
	\includegraphics[width=0.5\textwidth]{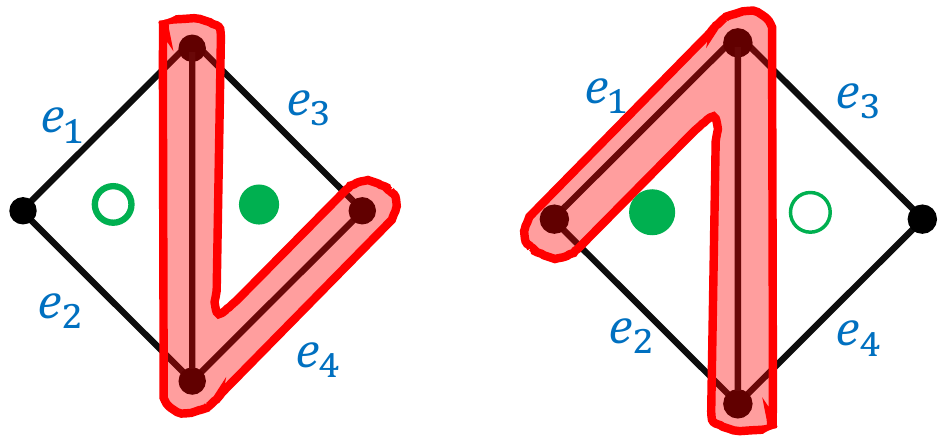}
	\caption{Edges (highlighted in red) that can be afflicted by an error from a single fault resulting in a left or right flag. A highlighted cross edge corresponds to $X$ or $Z$ data qubit errors on edges $e_1$ and $e_2$ (which is equivalent to two data qubit errors on edges labeled $e_3$ and $e_4$ up to a stabilizer). Such edges will be referred to as boomerang edges due to the shape of the highlighted area. }
	\label{fig:BoomerangEdges}
\end{figure}

In what follows, when a flag qubit has a non-trivial measurement outcome, we will say that the flag qubit flagged. We also assume the following depolarizing circuit level noise model :

\begin{enumerate}
        \item With probability $p$, each single-qubit gate location is followed by a Pauli error drawn uniformly and independently from $\{ X,Y,Z \}$.
	\item With probability $p$, each two-qubit gate is followed by a two-qubit Pauli error drawn uniformly and independently from $\{I,X,Y,Z\}^{\otimes 2}\setminus \{I\otimes I\}$.
	\item With probability $\frac{2p}{3}$, the preparation of the $\ket{0}$ state is replaced by $\ket{1}=X\ket{0}$. Similarly, with probability $\frac{2p}{3}$, the preparation of the $\ket{+}$ state is replaced by $\ket{-}=Z\ket{+}$.
	\item With probability $\frac{2p}{3}$, any single qubit measurement has its outcome flipped.
	\item Lastly, with probability $p$, each idle gate location is followed by a Pauli error drawn uniformly and independently from $\{ X,Y,Z \}$.
\end{enumerate}

When measuring the weight-four Pauli operators in \cref{fig:HeavyHexStabMeasurements,fig:HeavySquareXZStab}, we have already discussed how a single fault can lead to a weight-two data qubit error while at the same time resulting in a flag\footnote{Note that when measuring a weight-four $X$-type or $Z$-type operator, if a weight-two data qubit error occurs without a flag, at least two faults have occurred.} (see \cref{fig:FlagOutcomes}). In \cref{fig:HeavyHexGraphs,fig:MatchGraphHeavySquare}, we illustrated the matching graphs corresponding to $Z$ and $X$-type stabilizer measurements of the heavy hexagon and heavy square code by adding green vertices representing the flag measurement outcomes. For the heavy hexagon code, faults resulting in $X$ errors during the $X$-type gauge measurements can result in non-trivial flag outcomes, and the syndrome of the resulting data qubit errors is measured during the $Z$-type stabilizer measurements. Flag outcomes for the heavy square code have an analogous representation but are present for both $X$ and $Z$ stabilizer measurements. In \cref{fig:BoomerangEdges}, edges (data qubits) that can be afflicted by an error from a single fault resulting in a flag are shown. Due to the shape of the highlighted area, we will refer to such edges as boomerang edges. Each diamond has two green vertices (which we refer to as left or right flags) since two flag qubits are used to measure the weight-four operators of the heavy hexagon and heavy square code. Note that flags can also arise from measurement errors. Therefore flag qubits can flag without the presence of data qubit errors. However, by analyzing the circuits of \cref{fig:HeavyHexStabMeasurements,fig:HeavySquareXZStab}, it can be shown that a single fault which results in both left and right flags cannot induce data qubit errors. Thus when both left and right flags are highlighted, information from the flag qubit measurement outcomes is ignored.

The goal of this section is to present a new decoding algorithm which integrates the flag qubit measurement outcomes into the minimum weight perfect matching algorithm to ensure that errors arising from at most $(d-1)/2$ faults are always corrected\footnote{The flag methods presented in this section are currently being applied in other decoding schemes for topological codes such as the color code \cite{ColorCodeFlags}.}. The decoder should be efficient so that it can be implemented via Pauli frame updates \cite{Barbara15,CIP17}. In order to do so, we make the following observations. Note that a single fault resulting in a left or right flag occurs with probability $\mathcal{O}(p)$. In general, $m$ left or right flags, each arising from a single fault, will occur with probability $\mathcal{O}(p^m)$. Having both $m$ flags in addition to $l$ errors outside boomerangs is an $\mathcal{O}(p^{m+l})$ event.

\begin{figure}
	\centering
	\includegraphics[width=0.4\textwidth]{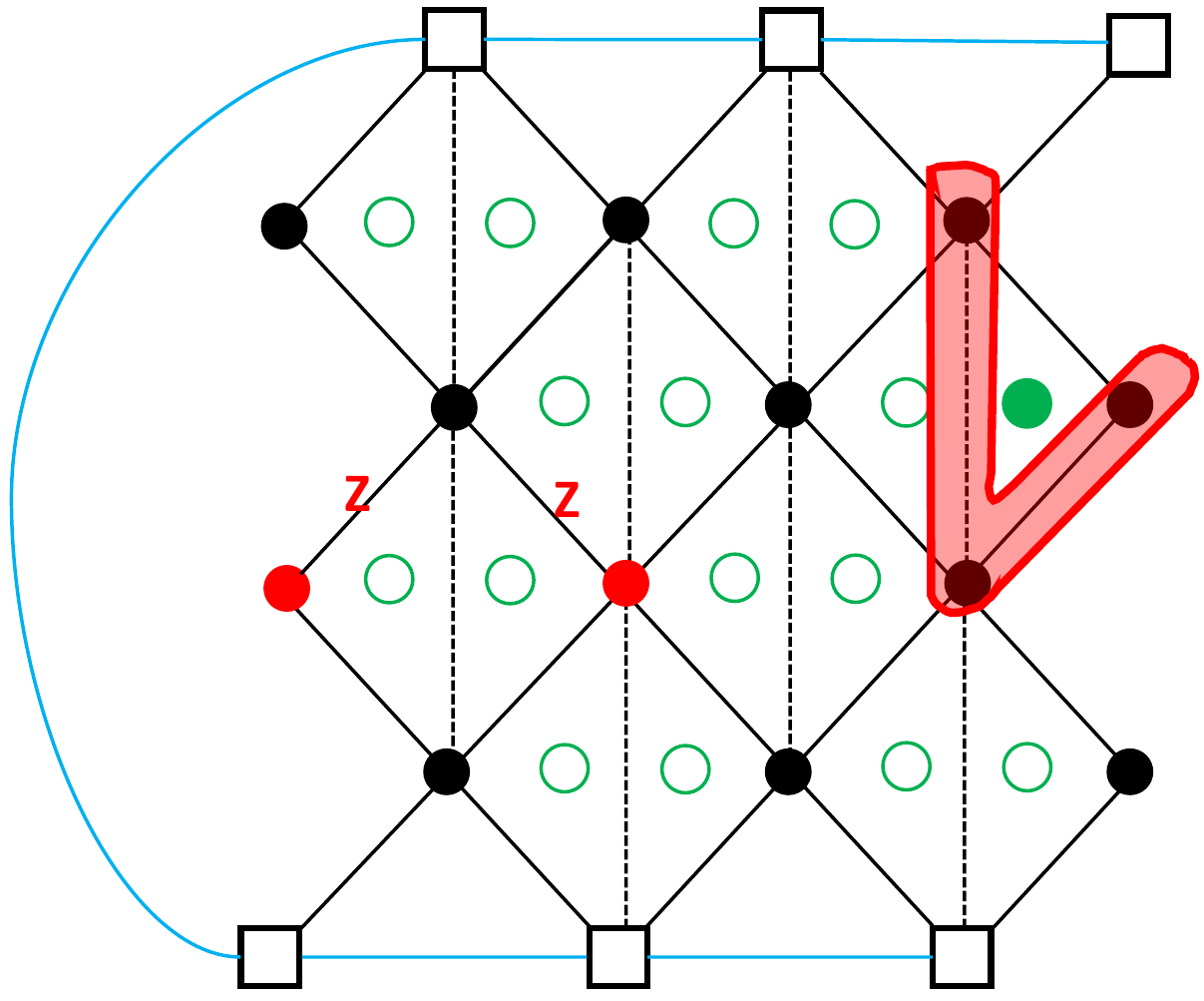}
	\caption{Example of two $Z$ errors resulting in two highlighted vertices for the graph $G_{X}$. In addition, we consider the case where another fault occurs which results in a flag as shown in the figure.}
	\label{fig:ExampleFlagTwoZ}
\end{figure}

Since data qubit errors within boomerang edges occur with probability $\mathcal{O}(p)$, weights of edges E outside of the boomerangs are renormalized so that $w_{\text{E}} = -\log{p^{m}P_{\text{E}}}$ whereas the weights of edges within the boomerangs are computed based on the leading order error configurations giving rise to those edges. More formally, the decoding protocol using the flag qubit measurement outcomes is given as follows:

\textbf{Decoding protocol using flag qubits:}
Consider a distance $d$ heavy hexagon or heavy square code. After performing $d$ rounds of error syndrome measurements, suppose there are a total of $m$ left or right flag outcomes associated with the graph $G$.

\begin{enumerate}
   \item Leave all edge weights inside the boomerangs corresponding to left or right flag outcomes unchanged.

   \item Let E be an edge outside a highlighted boomerang and $P_{\text{E}}$ the probability of all error configurations resulting in an error on edge E. Replace $P_{\text{E}}$ by  $P'_{\text{E}}=p^m P_{\text{E}}$.

   \item Replace the edge weight $w_{\text{E}}$ of E by $w'_{\text{E}} = -\log{P'_{\text{E}}}$.

   \item Define $G'$ to be the graph $G$ with new edge weights computed from the previous steps.

   \item Vertices in $G'$ are highlighted if the corresponding $X$- or $Z$-type stabilizer outcomes change in two consecutive rounds. If an odd number of vertices are highlighted, highlight a boundary vertex.

   \item Implement the minimum weight perfect matching algorithm on the graph $G'$ to identify all pairs of highlighted vertices to be matched.

\item Find the minimum weight path connecting all pairs of matched vertices of $G'$.

\item If $G'$ is a $d$-dimensional graph, the highlighted edges in $G'$ are mapped to edges in the corresponding $d-1$ dimensional planar graph added modulo two.

\item The correction is applied to the remaining highlighted edges.

\end{enumerate}

Note that the probabilities assigned to edges outside of the boomerang's do not always correspond to the correct probability distribution for such edges. As an example, suppose there are $m$ flags and two data qubit $Z$ errors outside the boomerang's for a graph associated with the $X$-stabilizer measurements. Assume that the $Z$ errors result in two highlighted vertices as for example, in \cref{fig:ExampleFlagTwoZ} (where one flag qubit flagged). The assigned probability for the path connecting the two vertices will be $\mathcal{O}(p^{2(m+1)})$ instead of the actual probability $\mathcal{O}(p^{m+2})$ and will thus have a higher weight. One could be concerned that the high preference to edges within a boomerang could distort paths such that a correctable error configuration (under standard minimum weight perfect matching) would go uncorrected. However as we show below, the decoder described above can correct error configurations arising from at most $\floor*{(d-1)/2}$ faults and is thus fault-tolerant.

\begin{figure*}
	\centering
	\includegraphics[width=0.8\textwidth]{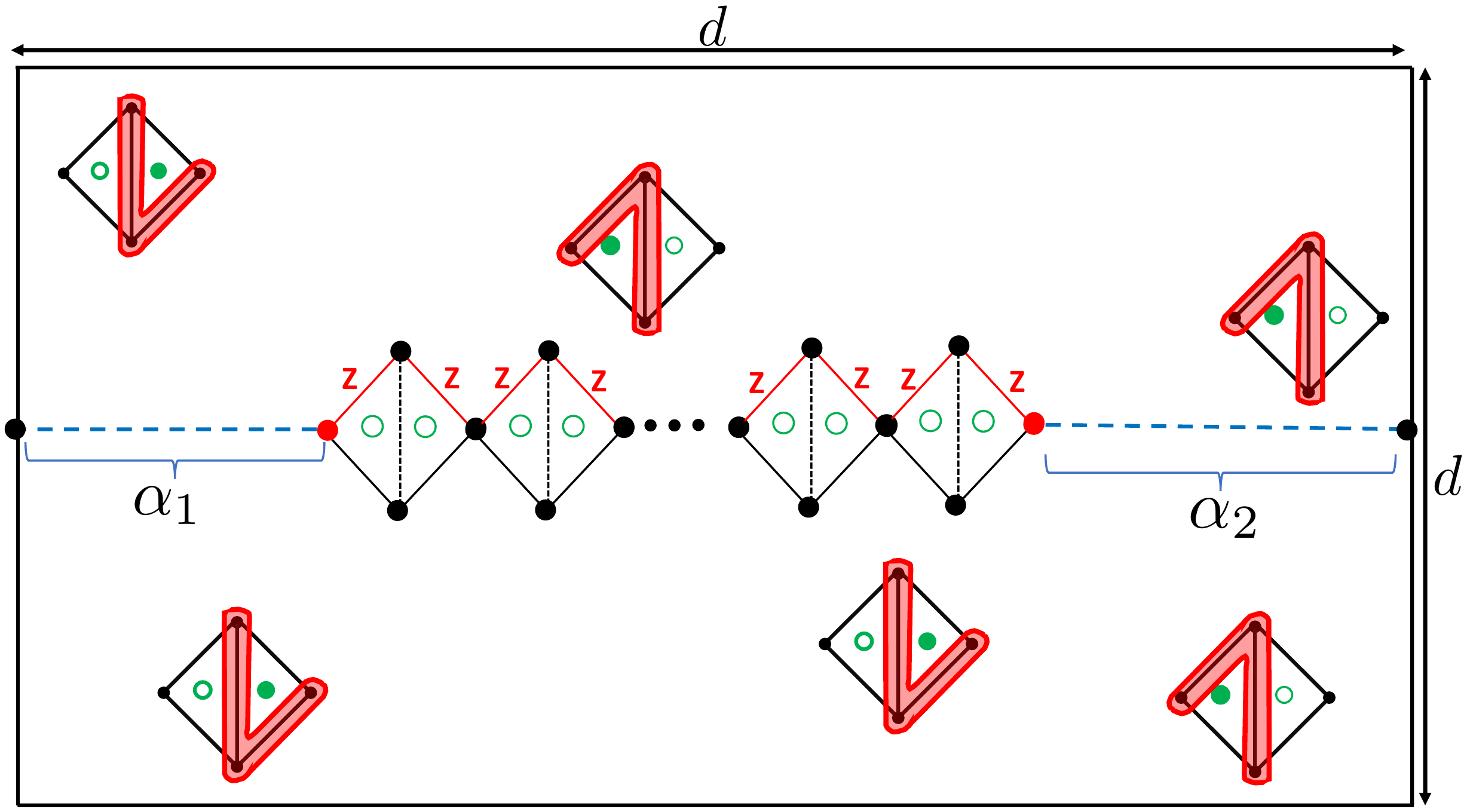}
	\caption{Illustration of a case where there are $m_1$ flags, each resulting from a measurement error, and $m_2$ consecutive $Z$ errors resulting (the particular error type is not important) in two highlighted vertices. We have the constraint that $m_1 + m_2 \le \floor*{(d-1)/2}$. The graph can be any graph associated to a distance $d$ heavy square or heavy hexagon code. Hence $\alpha_1 + \alpha_2 \ge \floor*{(d+1)/2}$. The dark lines represent the boundaries of the graph.}
	\label{fig:ProofFlagScheme}
\end{figure*}

Consider the worst case scenario where $m_1 > 0$ flags occur which are all caused by measurement errors so that paths within boomerangs contain no data qubit errors. In addition, suppose there are $m_2$ consecutive $X$ or $Z$ data qubit errors (whether it is $X$ or $Z$ is irrelevant as long as all errors are of the same type) which result in two highlighted vertices. We are interested in the case where $m_1 + m_2 \le \floor*{(d-1)/2}$ so that the total number of data qubit errors is correctable by the code. Thus the number of edges $\alpha_1$ and $\alpha_2$ connecting the two highlighted vertices to the nearest boundary of the graph satisfies $\alpha_1 + \alpha_2 \ge \floor*{(d+1)/2}$. An illustration is provided in \cref{fig:ProofFlagScheme}.

Clearly, the path which corrects all the data qubit errors is one which goes through all the diamonds shown in \cref{fig:ProofFlagScheme} which does not contain a boomerang (the edges belonging to the correct path are highlighted in red). However each edge $E$ along such path will have weight $w_{E} = -\log{p^{m_1}P_{E}}$ compared to the edges $E'$ in the boomerangs which will have weight $w_{E'} = -\log{P_{E'}}$. For the boomerangs to distort the minimum weight path connecting the highlighted vertices in such a way that a logical fault occurs, the path would need to connect the highlighted vertices to the boundary of the graph. But since $m_1 + m_2 \le \floor*{(d-1)/2}$, there must be at least $(d+1)/2$ edges along such a path that does not belong to a boomerang and thus has weight $w_{E} = -\log{p^{m_1}P_{E}}$. Consequently, such a path would have weight $w_1 \ge -\log{p^{m_1\floor*{(d+1)/2}}\prod_{E'}P_{E'}}$ compared to the path which corrects the errors which has weight $w_2 < -\log{p^{m_1\floor*{(d-1)/2}}\prod_{E}P_{E}}$ which has smaller weight\footnote{The products are over all edges $E'$ and $E$ along the diamonds of the paths of length at least $\floor*{(d+1)/2}$ and the path which corrects the errors which has length less than $\floor*{(d-1)/2}$.}. Therefore the minimum weight path will correct the errors as required.

\begin{figure}
	\centering
	\includegraphics[width=0.5\textwidth]{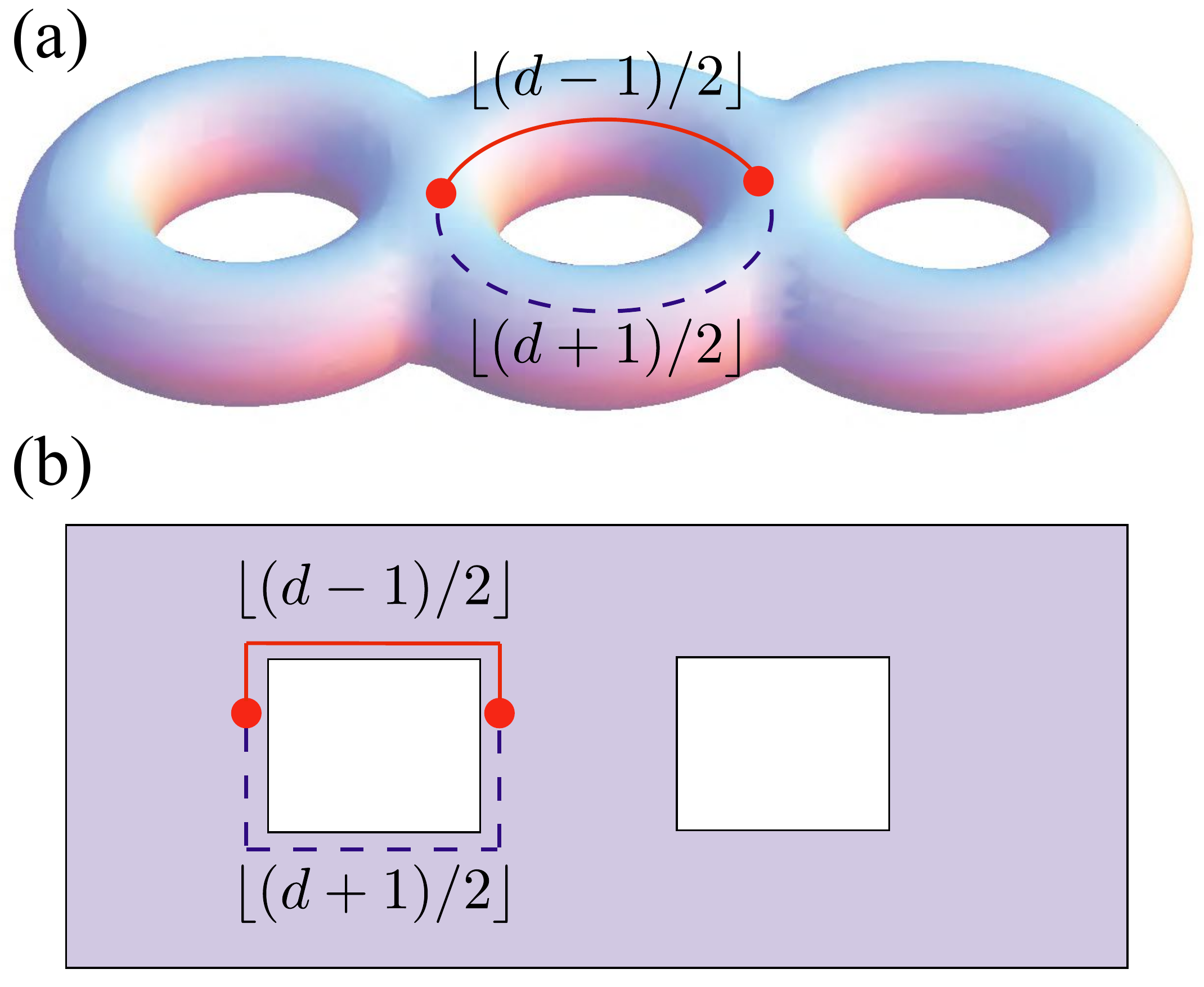}
	\caption{ Minimum-weight paths on (a) a genus-$g$ surface  and (b) a surface with hole defects. The red dots indicate the syndromes. The red lines correspond to the actual data qubit error string and the correct minimum-weight path. The blue dashed lines indicated the distorted path which will induce a logical error.}
	\label{fig:topology}
\end{figure}

We point out that in \cite{NeuralNetFlags}, information from flag qubit measurement outcomes were used in a neural network decoder to decode topological color codes resulting in improved thresholds. However the scheme is not scalable as it requires an exponential increase in training data as a function of the code distance.

Although the above discussion applies to heavy square/hexagon codes and more generally to topological stabilizer codes with open boundaries, it also applies to the cases when these codes are defined on a closed surface (no boundaries) with nonzero genus $g$. The above analysis can be straightforwardly adopted to the $g=1$ case,  i.e., codes defined on a torus, which can be constructed by identifying the opposite edges of  the square patch (periodic boundary condition) in \cref{fig:ProofFlagScheme}.  In the general genus-$g$ case,  the code distance $d$ is given by the systole of the entire surface, i.e., the shortest non-contractible loop.  The logical operators correspond to the non-contractible loops on the surface characterized by the first homology group $H_1 = \mathbb{Z}_2^{2g}$. In order to distort the minimum-weight path to form a non-contractible loop corresponding to a logical error, as illustrated by \cref{fig:topology}(a), the distorted path again needs to have at least $\floor*{(d+1)/2}$ edges outside boomerang edges which has higher weight than the actual error path with length $\floor*{(d-1)/2}$. 

Similarly, the same proof also  applies to the case where logical information is encoded using holes with gapped boundaries \cite{Cong:2017gza, FMMC12} \footnote{We expect that the case with twist defects \cite{bombin2010} also works. However, due to the introduction of different types of  stabilizers, our current scheme is incomplete for that purpose and we leave the detailed studies to future work. },  as illustrated by \cref{fig:topology}(b), as well as the most general case where all of these encoding schemes coexist \cite{Brown2017}.    We hence reach the following theorem:
 
 \begin{theorem}\label{theorem1}
 There exists topological stabilizer codes with flag qubits defined on a genus-$g$ surface with $p$ open boundaries, $q$ holes such that the flag decoding scheme achieves fault tolerance with the full code distance \\ 
 ($g$, $p$, $q \in \{0, \mathbb{N}^+\}$).	
 \end{theorem}
 
 \begin{figure*}
	\centering
	\subfloat[\label{fig:LogicalXHeavyHexv2}]{%
		\includegraphics[width=0.5\textwidth]{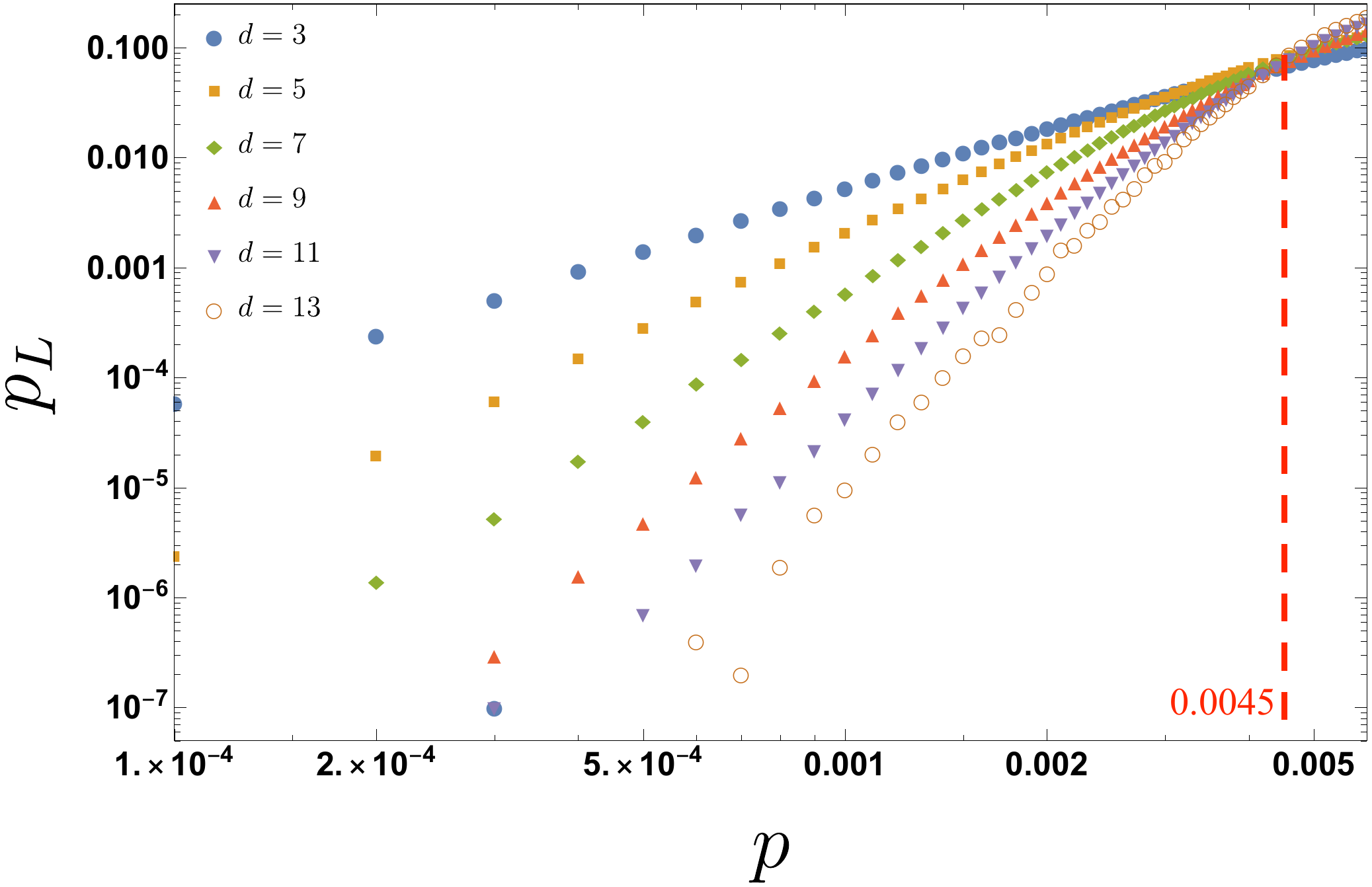}
	}
	\subfloat[\label{fig:LogicalZHeavyHexv2}]{%
		\includegraphics[width=0.5\textwidth]{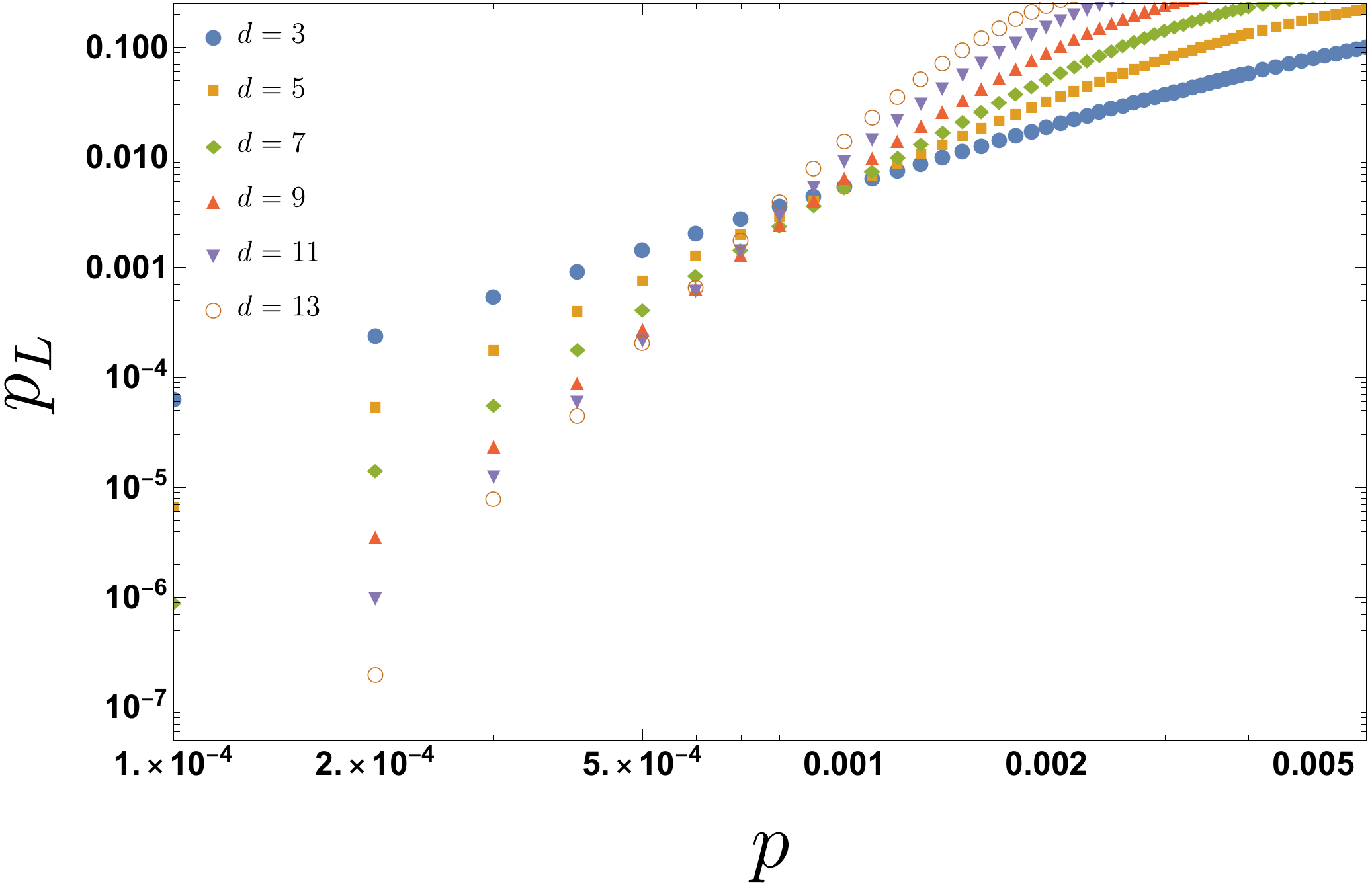}
	}
	\caption{(a) Logical $X$ error rates and (b) logical $Z$ error rates for the heavy hexagon code. The asymptotic threshold for logical $X$ errors is approximately $p_{\text{th}} = 0.0045$. Since $Z$ errors are corrected using Bacon-Shor type stabilizers, there is no threshold for $Z$ errors.}
	\label{fig:HeavyHexLogicalErrorRates}
\end{figure*}

\begin{figure*}
	\centering
	\subfloat[\label{fig:LogicalXHeavySquarev2}]{%
		\includegraphics[width=0.5\textwidth]{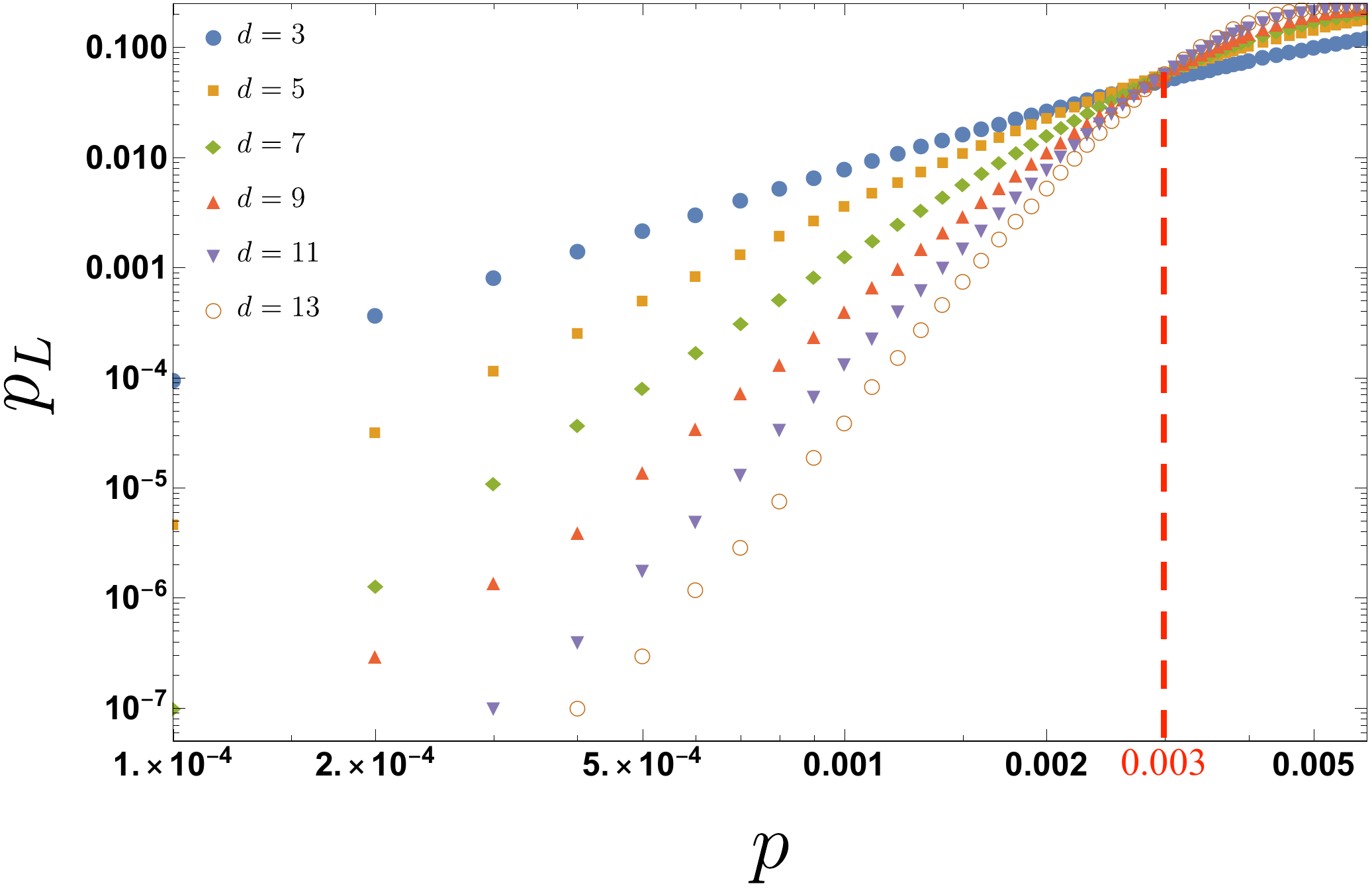}
	}
	\subfloat[\label{fig:LogicalZHeavySquarev2}]{%
		\includegraphics[width=0.5\textwidth]{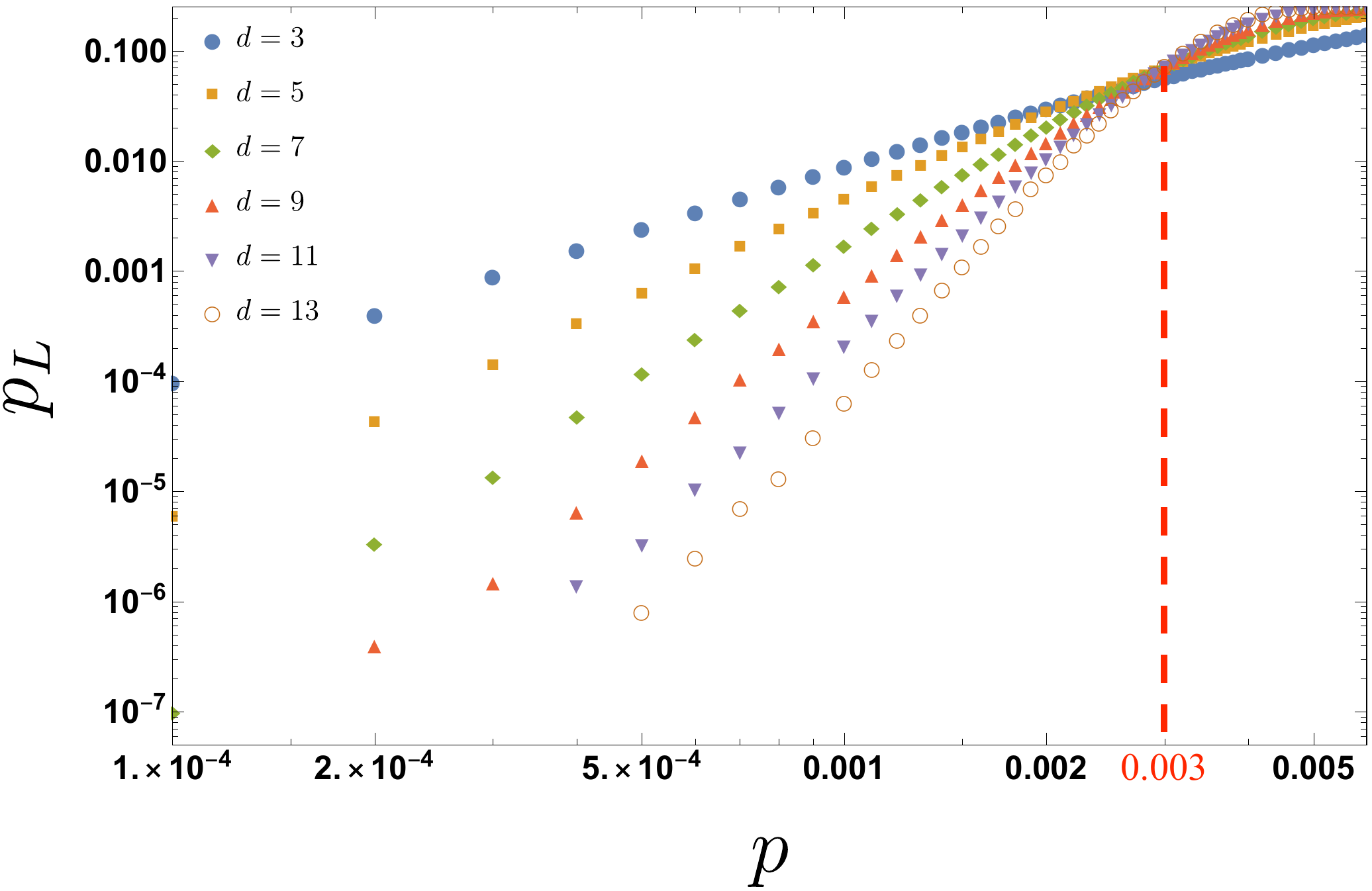}
	}
	\caption{(a) Logical $X$ error rates and (b) logical $Z$ error rates for the heavy square code. The asymptotic threshold can be seen to be approximately $p_{\text{th}} = 0.003$.}
	\label{fig:HeavySquareLogicalErrorRates}
\end{figure*}
 
 In this paper, we have proved the existence with the explicit construction of the heavy-square topological code, which has essentially the same topology-dependent logical encoding as the conventional toric code (e.g., both corresponding to the homology group $H_1 = \mathbb{Z}_2^{2g}$ on a genus-$g$ surface).  The explicit construction of (1) the heavy-square code on a high-genus surface by gluing two layers of punctured surfaces \cite{Zhu:2018CodeLong} and (2) the heavy-square code with hole defects \cite{Cong:2017gza, FMMC12}  are shown in \cref{app:genus}. We note that the methods described above apply to arbitrary constructions of high-genus surfaces with stabilizers of weight at most four, while the bilayer construction is the simplest one with local connectivity. We caution that more general constructions such as higher-genus hyperbolic surfaces will typically require stabilizer measurements of weight greater than four \cite{BTHype16,Breuckmann2017,Conrad18,LavasanZhu19} and thus the methods described above do not directly apply. However, with more flag qubits, it is possible to adapt the edge-weight renormalization scheme described in this work \cite{ColorCodeFlags}. The heavy-hexagon subsystem code, on the other hand,  is ``half-topological"  (corresponding to the $Z$-stabilizers) as mentioned above \footnote{We note that, for the heavy-hexagon code, the dependence of the logical subspace dimension and encoding on the topology is more subtle than an actual topological code, and will be discussed in future works.} and it turns out that the flag decoding scheme also achieves fault-tolerance with the full code distance. 
 
Lastly, we point out that for codes with stabilizer generators of weight $w>4$, depending on the support of logical operators, one could potentially require $v$-flag circuits to measure the stabilizer generators where $v = \frac{w}{2}-1$ (see \cite{CB17} for the definition of a $v$-flag circuit). If a $v$-flag circuit is required (with $v>1$), then edges should be renormalized based on the number of faults that resulted in a particular flag outcome. For instance, if two faults resulted in a particular flag outcome, then edges with support on the data qubits that could have errors resulting from the flags should be renormalized by $p$, and edges not in the support should be renormalized by $p^2$. Thus to guarantee the correction of errors arising from up to $\floor*{(d-1)/2}$ faults, extra flag qubits could be required to distinguish how many faults resulted in a particular flag outcome. 

\section{Numerical results}
\label{sec:Numerics}

\begin{figure*}
	\centering
	\subfloat[\label{fig:LogicXNoFlag}]{%
		\includegraphics[width=0.5\textwidth]{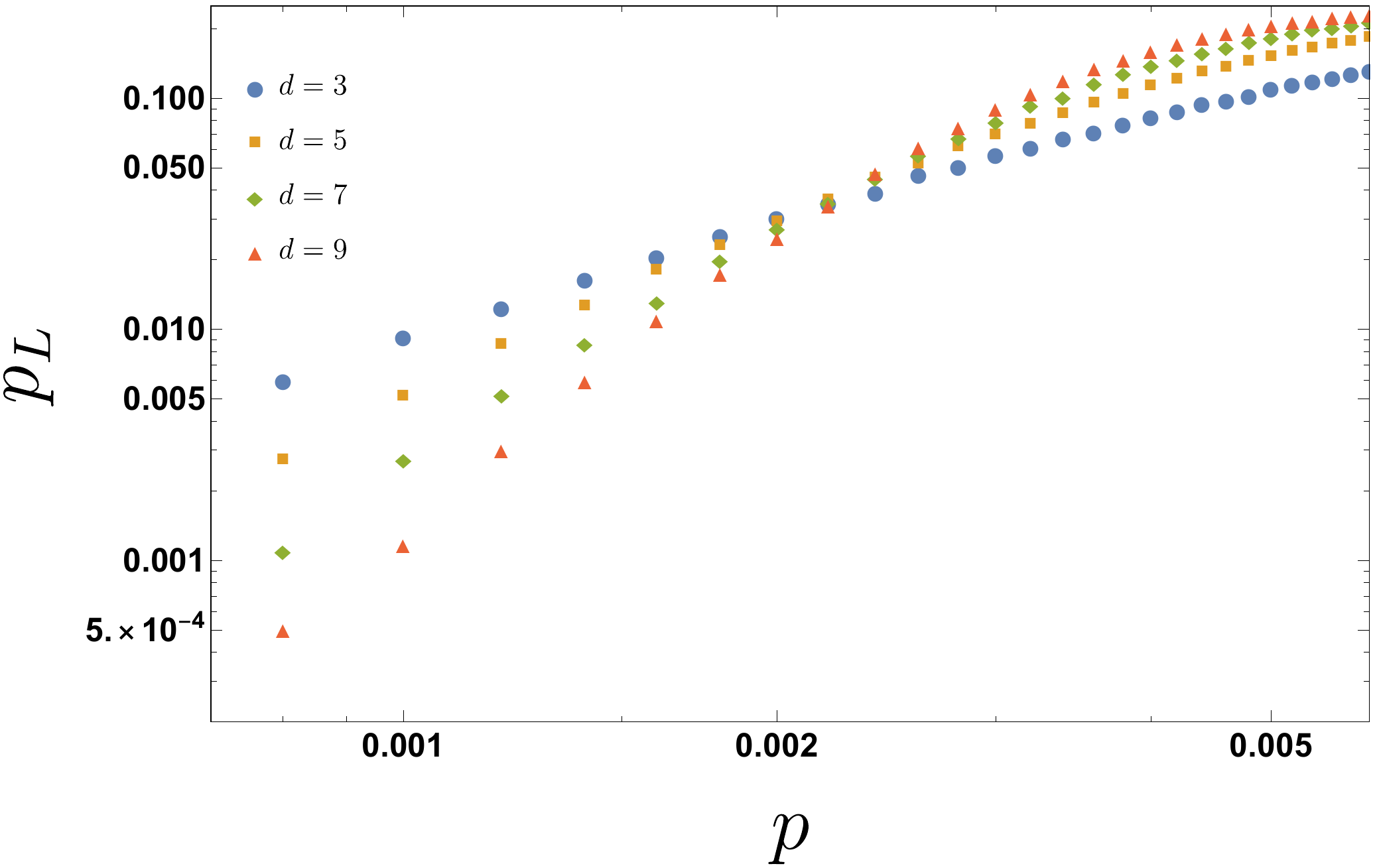}
	}
	\subfloat[\label{fig:LogicZNoFlag}]{%
		\includegraphics[width=0.5\textwidth]{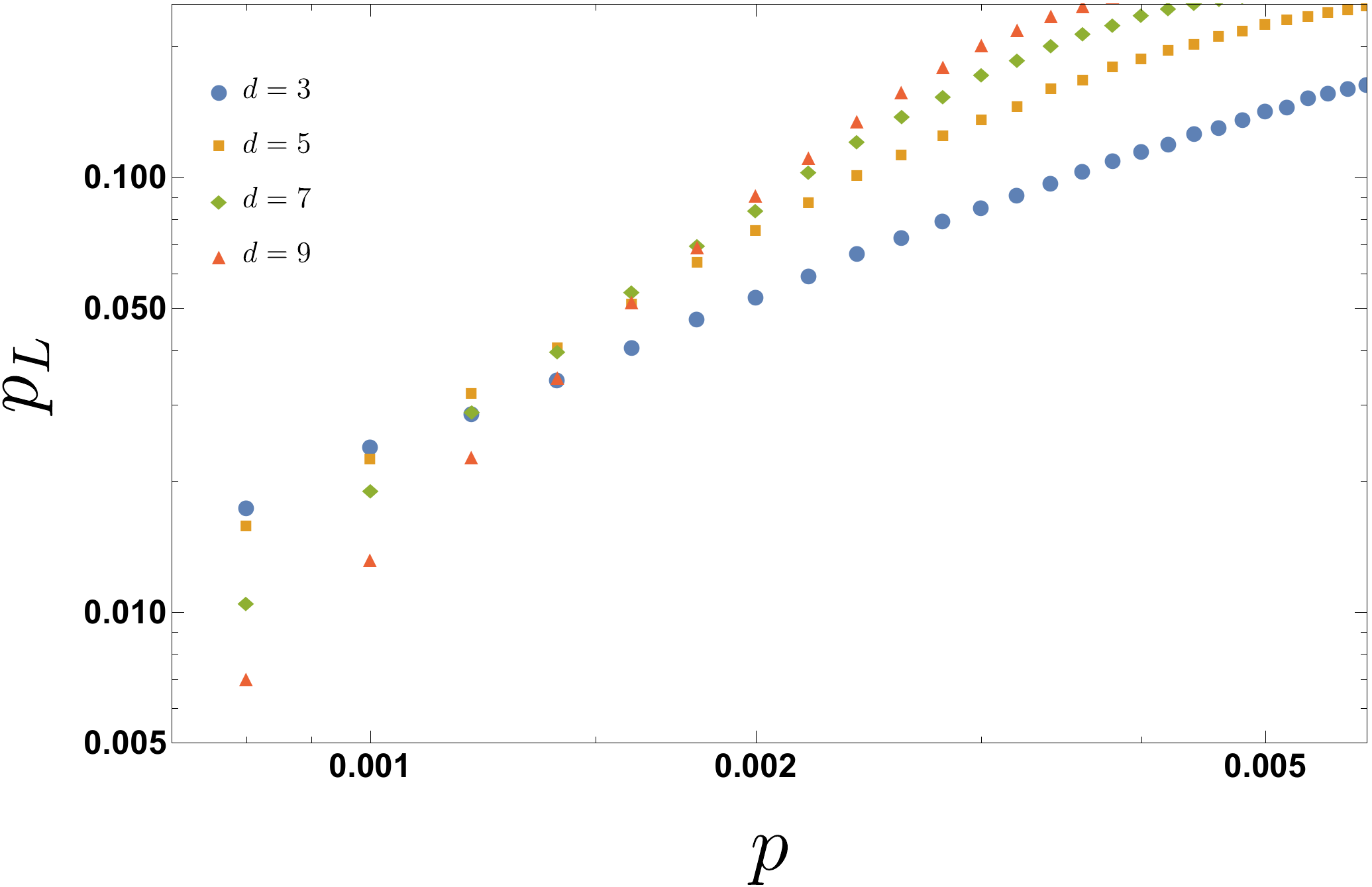}
	}
	\caption{(a) Logical $X$ error rates and (b) logical $Z$ error rates for the heavy square code when flag qubit information is ignored. The threshold for logical $X$ errors is approximately $p_{\text{th}} = 0.002$, half the value obtained when flag information is used to correct errors. Further, due to the error propagation properties of the heavy square code, logical $Z$ error rates are about an order of magnitude worse than compared to those obtained when flag information is kept.}
	\label{fig:LogicHeavyNoFlag}
\end{figure*}

Using the decoding protocol with flag qubits and the edge weights given in \cref{app:EdgeWeightCalc}, we computed the logical error rates of the heavy hexagon and heavy square codes for odd distances $3 \le d \le 13$. Error rates were computed by performing $10^7$ Monte Carlo simulations given the noise model described in \cref{sec:FlagSection}. Logical $X$ and $Z$ error rates\footnote{Note that if both a logical $X$ and $Z$ error occurs in a given Monte Carlo simulation, we count this as a logical $Y$ error, and logical $X$ and $Z$ errors are not incremented.} for both codes are given in \cref{fig:HeavyHexLogicalErrorRates,fig:HeavySquareLogicalErrorRates}.

For the heavy hexagon code, since $Z$ errors are corrected using Bacon-Shor type stabilizers, there is no threshold for logical $Z$ errors (see \cref{fig:LogicalZHeavyHexv2}). However for physical error rates close to $10^{-4}$, it can be seen that the logical error rate does decrease significantly for the code distances that were considered. $X$-type errors are corrected using a surface-code type decoding scheme. The $X$-error rate threshold was found to be $p_{\text{th}} = 0.0045$ which is fairly competitive with results obtained for the surface code \cite{BKM+14,FowlerMatching,FMMC12,FowlerEdgeWeights}.

Similarly to logical $X$ error rates of the heavy hexagon code, the heavy square code also exhibits high thresholds even though the circuit depth for the stabilizer measurements are 14 compared to 6 for the surface code. For both logical $X$ and $Z$ errors, the computed asymptotic threshold is found to be approximately $p_{\text{th}} = 0.003$.

A large reason for the high threshold values obtained (despite the large circuit depths) is due to our new decoding scheme which uses information from the flag qubit measurement outcomes. We already proved in \cref{sec:FlagSection} that the full code distance can be achieved. To further support our claims, we also computed the logical error rates for the heavy square code ignoring the flag qubit measurement outcomes and using standard minimum weight perfect matching methods used for the surface code. The plots are given in \cref{fig:LogicHeavyNoFlag}. As we discussed in \cref{subsec:heavysquare}, weight-two $Z$ errors arising from a single fault are parallel to the logical $Z$ operator of the heavy square code. It is thus not surprising to see that when flag information is ignored, the logical $Z$ error rates in \cref{fig:LogicZNoFlag} are about an order of magnitude higher than those in \cref{fig:LogicalZHeavySquarev2}. Further, the threshold for logical $X$ errors in \cref{fig:LogicXNoFlag} is approximately $p_{\text{th}} = 0.002$ which is less than half the value obtained when flag information is used to correct errors.

\section{Conclusion}
\label{sec:conclusion}

In this work we introduced two families of codes which we called the heavy hexagon code and heavy square code. The heavy hexagon code (where qubits are located on a heavy hexagonal lattice) is a hybrid surface/Bacon-Shor code where weight-four $X$-type gauge generators form Bacon-Shor type stabilizers and products of weight-two $Z$-type gauge generators form surface code type stabilizers. The heavy square code is a family of surface codes mapped onto a heavy square lattice. For superconducting qubit architectures using the CR gate, both code families achieve the goal of reducing the number of frequency collisions as compared to surface code devices. For a code of distance $d$, heavy square and heavy hexagon implementations require $10$ to $20\%$ more qubits than a rotated surface code. However when considering the practical effect of fabrication-related disorder $\sigma_f$ in the qubits' frequencies, for a given distance $d$ the heavy square and heavy hexagon codes achieve nearly an order of magnitude fewer frequency collisions than the rotated surface code, and can accept roughly twice the disorder while remaining collision-free.

One of the key ingredients in the fault-tolerant implementation of the above codes were the use of flag qubits for the weight-four $X$ and $Z$-type gauge and stabilizer measurements. We provided a scalable decoding scheme which makes use of the flag qubit information and can correct errors up to the full code distance. Performing Monte Carlo simulations for a depolarizing noise model, we showed that the heavy square code exhibits competitive threshold values (approximately $p_{\text{th}} = 0.003$) with the surface code. Since $Z$ errors are corrected via Bacon-Shor type decoding schemes for the heavy hexagon code, there is no threshold for $Z$ errors (although low logical error rates can be achieved for physical error rates in the order of $10^{-4}$). However the heavy hexagon code achieves a threshold of approximately  $p_{\text{th}} = 0.0045$ for $X$ errors.

In this work we also showed how our flag-qubit decoding scheme can be applied to codes defined on a surface with nonzero genus $g$. An interesting avenue for future work is to apply the flag-qubit decoding scheme to topological codes with stabilizer generators of weight greater than four (such as the color code) to ensure that errors are corrected up to the full code distance. Such an approach could result in thresholds which are closer to thresholds obtained for surface codes. When including the overhead cost of performing a universal gate set, such codes with flag qubit decoders could be a preferred alternative to the surface code. Another interesting avenue would be to extend the ideas presented in this work to topological codes with twist defects \cite{bombin2010} and to more general subsystem codes. We also point out that in the presence of $m$ flags, instead of renormalizing edge probabilities for edges outside boomerangs by $p^{m}P_{E}$, numerical optimizations could be performed to find the optimal coefficient $\alpha$ (potentially using machine learning techniques) such that edge probabilities would be renormalized by $p^{\alpha}P_{E}$. Such optimizations will inevitably be highly dependent on the underlying noise model afflicting the stabilizer measurement circuits.

Lastly, in Ref.~\cite{KenBrownLeakage} it was shown how some families of subsystem codes achieve better error correcting capabilities compared to the surface code in the presence of leakage errors. An interesting direction for future work would be to analyze the performance of codes defined on low degree graphs in the presence of leakage errors to see if such codes also have favorable error correcting capabilities compared to more standard implementations such as in the surface code.

\section{Acknowledgements}

We thank Jay Gambetta and Jerry Chow for useful discussions and for bringing the fabrication difficulties related to frequency collisions to our attention. We thank Easwar Magesan for explaining the conditions that lead to frequency collisions. We thank John Smolin for discussions and the help with parallel computing. We also thank Mike Newman and Ken Brown for useful discussions.

\appendix

\section{Edge weights calculations for the matching graphs}
\label{app:EdgeWeightCalc}

\begin{figure*}
	\centering
	\includegraphics[width=0.99\textwidth]{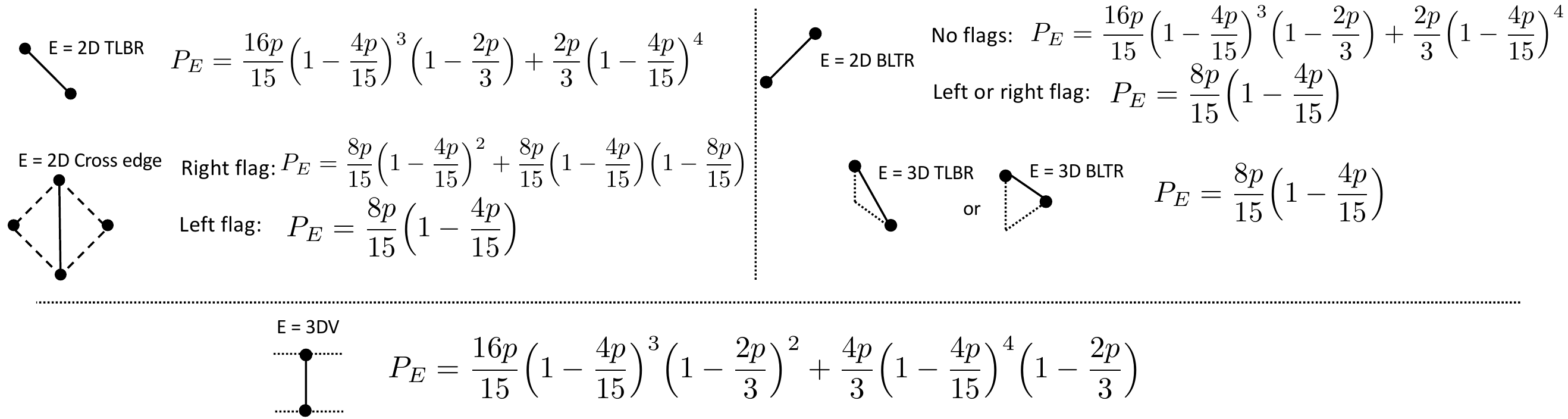}
	\caption{Edges for the graph obtained from $Z$-type stabilizer measurements of the heavy-hexagon code (\cref{fig:ZHexGraph,fig:ZHexGraphWithDiag}) and their associated probabilities to leading order in $p$. The label TLBR should be read as top left to bottom right. Similarly, the label BLTR should be read bottom left to top right. The cross edge is only activated when a left or right flag occurs. Lastly, 3DV is the vertical edge associated with measurement errors in the syndrome measurement qubits qubits.}
	\label{fig:HeavyHexZstabEdgeWeights}
\end{figure*}

In this appendix we provide examples of how to compute the weights of the edges for the graphs of \cref{fig:HeavyHexGraphs,fig:MatchGraphHeavySquare}. We then give the edge weights for all edges in the graphs used for correcting $X$ and $Z$ Pauli errors.

Consider the circuit containing the scheduling of the CNOT gates of the heavy hexagon code in  \cref{fig:HeavyHexCircuit}. In what follows, we focus on CNOT gates in the bulk of the lattice. An error of the form $XX$ occurring after the CNOT gate implemented at time step 8 for a $Z$-type parity measurement will result in a $X$ error on the corresponding data qubit, which will then propagate through the CNOT gate implemented in the ninth time step. Thus both $Z$-type parity measurements interacting with a bulk data qubit will detect the $X$ error in one measurement round and will contribute to the edge weight of 2D TLBR (see \cref{fig:HeavyHexZstabEdgeWeights}). The full list of errors which contribute to the weight of 2D TLBR are $\{ XX, YY, XY, YX \}$ for CNOT's at the eight time step, $\{ XI, XZ, YI, YZ \}$ for CNOT's at the ninth time step (on the lower right of a white face), $\{ IX, ZX, IY, ZY \}$ for CNOT's at the third time step and fourth time step  (lower left of the $X$-type gauge measurement) and data qubit $X$ and $Y$ errors. From the noise model defined in \cref{sec:FlagSection}, the total probability (to leading order in $p$) for an error to result in an edge $E$ of type 2D TLBR is given by $P_{E} = \frac{16p}{15} (1-\frac{4p}{15})^3(1-\frac{2p}{3}) + \frac{2p}{3}(1-\frac{4p}{15})^4$.

On the other hand, an error of type $XI$ occurring after a CNOT at the eight time step (where $X$ is on the control qubit of the CNOT) will introduce a data qubit error which will propagate through the CNOT applied at the ninth time step. Hence for the measurement round at which the error occurred, only one of the two syndrome measurement qubits in the $Z$-type parity measurements will be highlighted. In the next measurement round, the data qubit will have an $X$ error so that both of the syndrome measurement qubits of the $Z$-type parity measurement will be highlighted. Therefore, such an error will result in the edge 3D TLBR of \cref{fig:HeavyHexZstabEdgeWeights}. The types of errors leading to such an edge are $\{ XI, XZ, YI, YZ \}$ for CNOT's at the eight time step and $\{ XX, YY, XY, YX \}$ for CNOT's at the ninth time step (bottom right of a white face in \cref{fig:HeavyHexCircuit}). Hence to leading order in $p$, the probability associated with the edge 3D TLBR is given by $P_{E} = \frac{8p}{15} (1- \frac{4p}{15})$. The probabilities associated with the other edges of \cref{fig:HeavyHexZstabEdgeWeights} can be computed using similar methods as the ones described above.

Using the same methods as above, we computed the edge weights for the edges corresponding to the $X$-type gauge measurements of the heavy-hexagon code (\cref{fig:BaconGraph}). However one important difference is that only the odd parity of a configuration of errors are relevant when decoding Bacon-Shor type codes. Taking into account all odd configurations of errors giving rise to particular edge types, we obtain
\begin{widetext}

\begin{align}
\nonumber P_{b_1}=&\sum_{n=1}^{\frac{d+1}{2}}\sum_{m=0}^{\frac{3d-3}{2}}\left(^d_{2n-1}\right)\left(\frac{2p}{3}\right)^{2n-1}\left(1-\frac{2p}{3}\right)^{d-(2n-1)}\left(^{3d-2}_{2m}\right) \left(\frac{8p}{15}\right)^{2m}\left(1-\frac{8p}{15}\right)^{3d-2-2m} \\
&+ \sum_{n=0}^{\frac{d-1}{2}}\sum_{m=1}^{\frac{3d-1}{2}}\left(^d_{2n}\right)\left(\frac{2p}{3}\right)^{2n}\left(1-\frac{2p}{3}\right)^{d-2n}\left(^{3d-2}_{2m-1}\right) \left(\frac{8p}{15}\right)^{2m-1}\left(1-\frac{8p}{15}\right)^{3d-2-(2m-1)},
\end{align}
\begin{align}
P_{d_2}=\frac{1}{2}-\frac{1}{2}\left(\frac{8p}{15}-1\right)^{2d-2},
\end{align}
\begin{align}
P_{d_1}=\frac{1}{2}-\frac{1}{2}\left(1-\frac{8p}{15}\right)^{2d},
\end{align}
\begin{align}
P_{d'_1}=\frac{8p}{15}\left(1-\frac{4p}{15}\right),
\end{align}
\begin{align}
\nonumber P_{b_u}=&\sum_{n=1}^{\frac{d+1}{2}}\sum_{m=0}^{d-1}\sum_{l=0}^{d}\left(^{d}_{2n-1}\right)\left(\frac{2p}{3}\right)^{2n-1}\left(1-\frac{2p}{3}\right)^{d-(2n-1)}\left(^{2(d-1)}_{2m}\right) \left(\frac{8p}{15}\right)^{2m}\left(1-\frac{8p}{15}\right)^{2(d-1)-2m} \\
\nonumber & \cdot \left(^{2d}_{2l}\right) \left(\frac{4p}{15}\right)^{2l}\left(1-\frac{4p}{15}\right)^{2d-2l} + \sum_{n=0}^{\frac{d-1}{2}}\sum_{m=1}^{d-1}\sum_{l=0}^{d}\left(^{d}_{2n}\right)\left(\frac{2p}{3}\right)^{2n}\left(1-\frac{2p}{3}\right)^{d-2n} \\
\nonumber & \cdot \left(^{2(d-1)}_{2m-1}\right) \left(\frac{8p}{15}\right)^{2m-1}\left(1-\frac{8p}{15}\right)^{2(d-1)-(2m-1)}\left(^{2d}_{2l}\right) \left(\frac{4p}{15}\right)^{2l}\left(1-\frac{4p}{15}\right)^{2d-2l} \\
\nonumber &+\sum_{n=0}^{\frac{d-1}{2}}\sum_{m=0}^{d-1}\sum_{l=1}^{d}\left(^{d}_{2n}\right)\left(\frac{2p}{3}\right)^{2n}\left(1-\frac{2p}{3}\right)^{d-2n}\left(^{2(d-1)}_{2m}\right) \left(\frac{8p}{15}\right)^{2m}\left(1-\frac{8p}{15}\right)^{2(d-1)-2m} \\
\nonumber &\cdot \left(^{2d}_{2l-1}\right) \left(\frac{4p}{15}\right)^{2l-1}\left(1-\frac{4p}{15}\right)^{2d-(2l-1)} +\sum_{n=1}^{\frac{d+1}{2}}\sum_{m=1}^{d-1}\sum_{l=1}^{d}\left(^{d}_{2n-1}\right)\left(\frac{2p}{3}\right)^{2n-1}\left(1-\frac{2p}{3}\right)^{d-(2n-1)}\left(^{2(d-1)}_{2m-1}\right)\\
& \cdot \left(\frac{8p}{15}\right)^{2m-1}\left(1-\frac{8p}{15}\right)^{2(d-1)-(2m-1)}\left(^{2d}_{2l-1}\right) \left(\frac{4p}{15}\right)^{2l-1}\left(1-\frac{4p}{15}\right)^{2d-(2l-1)},
\end{align}
\begin{align}
\nonumber P_{m}=&\sum_{n=1}^{\frac{d+1}{2}}\sum_{m=0}^{\frac{d-1}{2}}\sum_{l=0}^{\frac{d-1}{2}}  \left(^{d+1}_{2n-1}\right)\left(\frac{4p}{15}\right)^{2n-1}\left(1-\frac{4p}{15}\right)^{d+1-(2n-1)}\left(^{d-1}_{2m}\right) \left(\frac{8p}{15}\right)^{2m}\left(1-\frac{8p}{15}\right)^{d-1-2m}   \\
\nonumber & \left(^{d+1}_{2l}\right)\left(\frac{2p}{3}\right)^{2l}\left(1-\frac{2p}{3}\right)^{d-2l}   + \sum_{n=0}^{\frac{d+1}{2}}\sum_{m=1}^{\frac{d-1}{2}}\sum_{l=0}^{\frac{d-1}{2}}  \left(^{d+1}_{2n}\right)\left(\frac{4p}{15}\right)^{2n}\left(1-\frac{4p}{15}\right)^{d+1-2n} \\
\nonumber & \left(^{d-1}_{2m-1}\right) \left(\frac{8p}{15}\right)^{2m-1}\left(1-\frac{8p}{15}\right)^{d-1-(2m-1)}\left(^{d+1}_{2l}\right)\left(\frac{2p}{3}\right)^{2l}\left(1-\frac{2p}{3}\right)^{d-2l} \\
\nonumber & +\sum_{n=0}^{\frac{d+1}{2}}\sum_{m=0}^{\frac{d-1}{2}}\sum_{l=1}^{\frac{d+1}{2}}  \left(^{d+1}_{2n}\right)\left(\frac{4p}{15}\right)^{2n}\left(1-\frac{4p}{15}\right)^{d+1-2n}\left(^{d-1}_{2m}\right) \left(\frac{8p}{15}\right)^{2m}\left(1-\frac{8p}{15}\right)^{d-1-2m}   \\
\nonumber & \left(^{d+1}_{2l-1}\right)\left(\frac{2p}{3}\right)^{2l-1}\left(1-\frac{2p}{3}\right)^{d-(2l-1)} + \sum_{n=1}^{\frac{d+1}{2}}\sum_{m=1}^{\frac{d-1}{2}}\sum_{l=1}^{\frac{d+1}{2}}  \left(^{d+1}_{2n-1}\right)\left(\frac{4p}{15}\right)^{2n-1}\left(1-\frac{4p}{15}\right)^{d+1-(2n-1)}\\
& \left(^{d-1}_{2m-1}\right) \left(\frac{8p}{15}\right)^{2m-1}\left(1-\frac{8p}{15}\right)^{d-1-(2m-1)}\left(^{d+1}_{2l-1}\right)\left(\frac{2p}{3}\right)^{2l}\left(1-\frac{2p}{3}\right)^{d-(2l-1)}.
\end{align}

\end{widetext}

Lastly, the probabilities associated with the edges for $X$ and $Z$ stabilizer measurements of the heavy square code are the same as those in \cref{fig:HeavyHexZstabEdgeWeights} except for the edge 3DV, which has $P_{E} = \frac{32p}{15} ( 1- \frac{8p}{15})^3 (1-\frac{4p}{15})^4 (1- \frac{2p}{3}) + \frac{16p}{15} ( 1- \frac{8p}{15})^4 (1-\frac{4p}{15})^3 (1- \frac{2p}{3}) + \frac{2p}{3} ( 1- \frac{8p}{15})^4 (1-\frac{4p}{15})^4$ (for both graphs in \cref{fig:MatchGraphHeavySquare})

\section{Encoding logical qubits into a high-genus surface and a surface with hole defects}
\label{app:genus}

In this appendix, we explicitly construct the heavy-square codes on a high-genus surface, or a surface with hole defects \cite{BK98, FM98}, in order to prove Theorem \ref{theorem1}.  In this way, we also show the more general encoding scheme that multiple logical qubits are encoded into a single code block, which can facilitate the logical gate operation.  

\begin{figure}
	\centering
	\includegraphics[width=1\columnwidth]{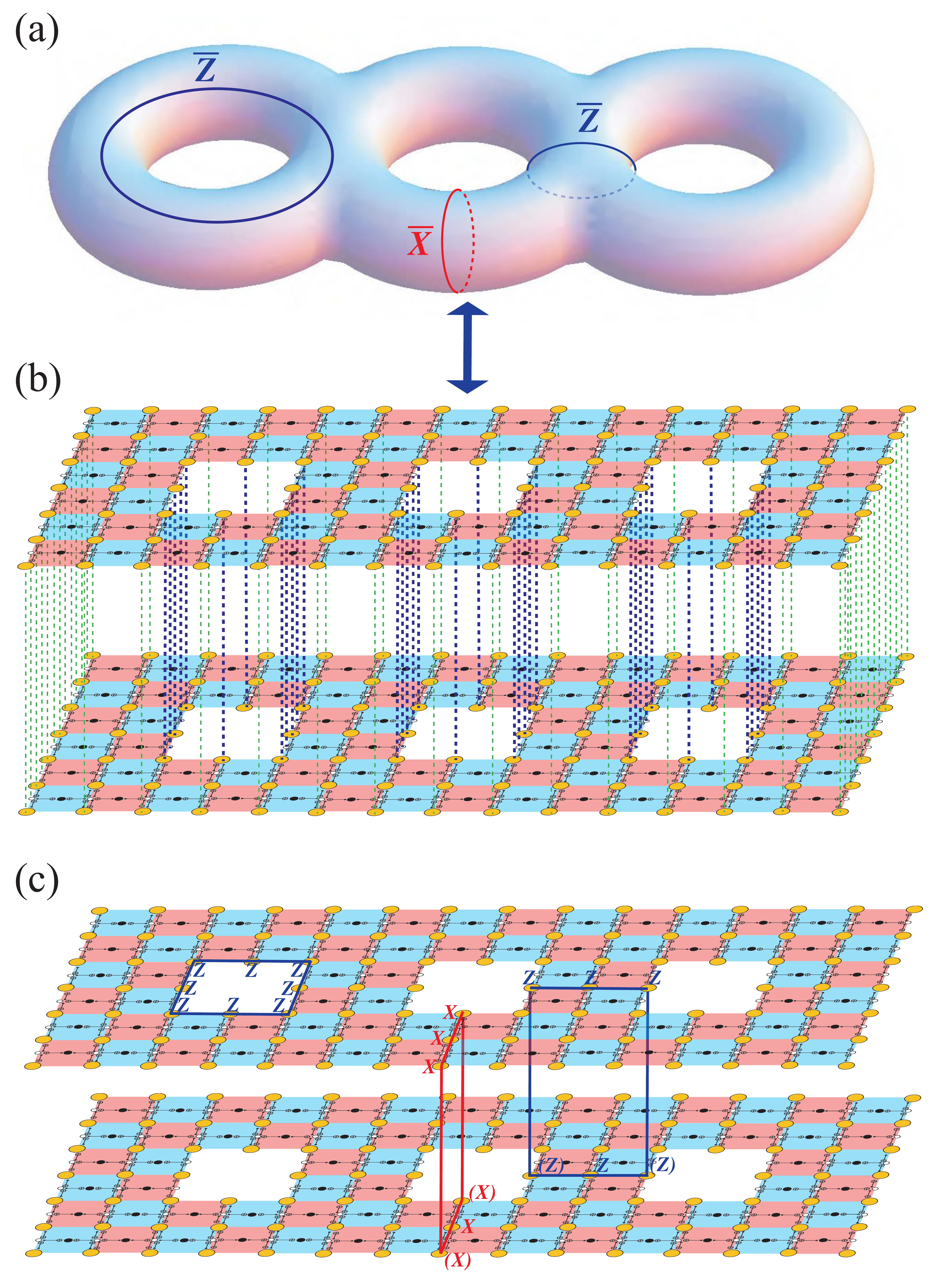}
	\caption{(a) A $g=3$ surface with the illustration of logical strings on three non-contractible cycles. (b) Explicit construction of the $g=3$ surface by identifying the hole boundaries and outer edges of two layers of heavy-square topological codes.   The blue and green dashed lines show the identification of the hole boundaries and outer edges respectively.  The vertically aligned plaquettes on the top and bottom layers have different types of stabilizers (indicated by different colors). (c) Logical strings corresponding to those shown in panel (a).  The parenthesis indicate Pauli operators identified with those on the top layers, which hence should not be included when counting the weight of the logical strings.}
	\label{fig:high_genus}
\end{figure}

\begin{figure}
	\centering
	\includegraphics[width=1\columnwidth]{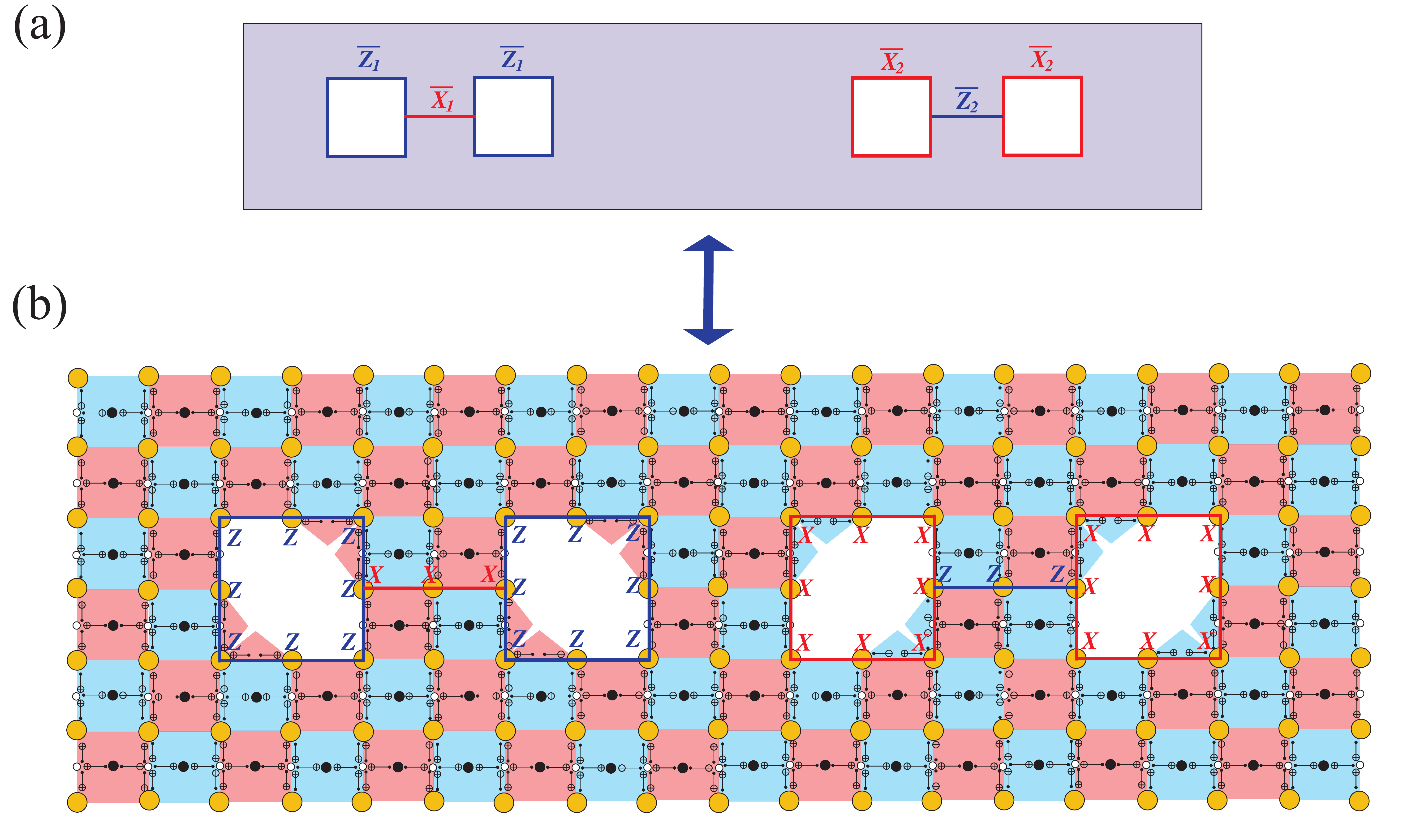}
	\caption{(a). On the left, there is a double Z-cut logical qubit consisting of two smooth whole defects; on the right, there is a double X-cut logical qubit consisting of two rough hole defects. (b) The explicit construction of the two types of logical qubits with hole defects on the heavy-square topological code.}
	\label{fig:holes}
\end{figure}

\subsection{High-genus surface}
\label{subsec:HighGenus}

A straightforward way to construct a high-genus surface is to have two layers of heavy-hexagon codes with holes, and ``glue" them together along the boundaries of the holes and the outer edges of the two layers \cite{Zhu:2018CodeLong}. Here, by ``gluing", we mean identifying the corresponding qubits along the top and bottom layers along the corresponding boundaries. In other words, the two identified qubits on the two layers are experimentally implemented with a single qubit.  We illustrate the construction of a $g=3$ surface in \cref{fig:high_genus}(a,b).  Here, the vertical dashed lines in \cref{fig:high_genus}(b) show all the identified data and syndrome measurement qubits. In particular, the blue dashed lines represent the gluing of the the hole boundaries and the green dashed lines represent the gluing of the outer edges of the two layers. Note that the X (red) and Z (blue) plaquettes aligned vertically are switched for the top and bottom layers, which makes sure that the neighbors of the X (red) plaquettes are always the Z (blue) plaquettes and vice versa.   

We also illustrate the corresponding logical strings in the bilayer systems in \cref{fig:high_genus}(c) which corresponds to the logical strings along three different non-contractible cycles in \cref{fig:high_genus}(a). As we can see these three logical strings have minimum operator support  6, 4, and 4 respectively.  Therefore, in this specific example, the length of the systole (i.e. the shortest non-contractible loop) of this surface and hence the code distance is $d=4$.  Typically one will construct the surface such that all these logical strings have the same length in order to optimize the information storage. As we have stated in Theorem \ref{theorem1}  in Sec.~\ref{sec:FlagSection}, the flag decoder can correct up to $\floor*{(d-1)/2}$ faults,  i.e., with weight less than half of the systole length. This can be seen from the fact that the measured operators have weight at most four. Therefore weight-two errors arising from a single fault will lie in boomerang edges, and the same arguments as in \cref{sec:FlagSection} apply.

We note that a bi-layer architecture for superconducting qubit lattices is within the reach of current superconducting quantum computing technology \cite{Gambetta2017}.   The inter-layer coupling can be implemented via 3D integration technologies such as the flip-chip \cite{Rosenberg2017} and the bump-bond \cite{Foxen_2017} architectures, as well as the thru-substrate vias (TSVs) architecture \cite{Vahidpour2017}.  The stabilizers near the edge of glued punctures (which we call \textit{handles}) will necessary need vertical inter-layer couplings with the 3D integration technologies.

All possible code deformations correspond to the mapping class group of the genus-$g$ surface, which can be generated by three types of Dehn twists ($3g-1$ in total) \cite{Zhu:2018CodeLong}.  The Dehn twists can be implemented in $O(1)$ time via a constant-depth circuit (independent of the code distance $d$) using long-range connectivity according to the protocol in Ref.~\cite{Zhu:2018CodeLong}.  With only local connectivity, one can implement the Dehn twists in $O(d)$ time with the protocol in Ref.~\cite{Koenig:2010do}. The protocol in Ref.~\cite{Koenig:2010do} is formulated for the general situation of Turaev-Viro codes, which contains the surface code as a specific case. Finally, we note that since the heavy-square code has the same braiding statistics as the surface code, the code deformation corresponding to the mapping class group of the genus-$g$ surface is contained in the Clifford group.

\subsection{Hole defects}
\label{subsec:HoleDefectsApp}

A more experimentally feasible way is to encode multiple logical qubits into a single-layer planar code with hole defects \cite{Cong:2017gza, FMMC12}.  As shown in \cref{fig:holes},  we can construct two types of hole defects equivalent to the smooth defect (Z-cut qubit) and rough defect (X-cut qubit)  in the standard surface codes respectively \cite{FMMC12}.  As shown in  \cref{fig:holes}(b), the smooth defect (Z-cut) has only X-type weight-two stabilizers on the hole boundary, while the rough defect (X-cut) has only Z-type weight-two stabilizers on the hole boundary.   In this example, we show a ``double Z-cut qubit" on the left and a   ``double X-cut qubit" on the right.  The double Z-cut qubit has the logical Z-string going around the holes, while the logical X-string connecting the two wholes, with length 6 and 3 respectively.  On the other hand, the double X-cut qubit has the logical X-string going around the holes, while the logical X-string connecting the two wholes.  The code distance in this case is $d=3$, i.e. the length of the shortest logical string.    Again, the flag decoder can correct up to $\floor*{(d-1)/2}$ faults in this case as well.  Braiding of a Z-cut defect around another X-cut defect implements a logical CNOT gate.

\bibliographystyle{unsrtnat} 
\bibliography{bibtex_chamberland.bib}

\end{document}